\documentclass[%
 reprint,
superscriptaddress,
noeprint,
 amsmath,amssymb,
 prl,
]{revtex4-2}
\usepackage{gensymb}
\usepackage{graphicx}
\usepackage{dcolumn}
\usepackage{xcolor}
\usepackage{bm}
\usepackage[hidelinks]{hyperref}

\newcommand{\siv}{SiV$^{0}$}
\newcommand{\sivm}{SiV$^-$}

\begin{document}


\title{Neutral silicon vacancy centers in undoped diamond via surface control}

\author{Zi-Huai Zhang}
\affiliation{%
Department of Electrical and Computer Engineering, Princeton University, Princeton, New Jersey 08544, USA
}%

\author{Josh A. Zuber}
\affiliation{
Department of Physics, University of Basel, Klingelbergstrasse 82, 4056 Basel, Switzerland
}
\affiliation{Swiss Nanoscience Institute, Klingelbergstrasse 82, 4056 Basel, Switzerland}

\author{Lila V. H. Rodgers}
\affiliation{%
Department of Electrical and Computer Engineering, Princeton University, Princeton, New Jersey 08544, USA
}%

\author{Xin Gui}
\affiliation{
Department of Chemistry, Princeton University, Princeton, New Jersey 08544, USA
}

\author{Paul Stevenson}
\affiliation{
Department of Physics, Northeastern University, Boston, Massachusetts 02115, USA
}

\author{Minghao Li}
\affiliation{
Department of Physics, University of Basel, Klingelbergstrasse 82, 4056 Basel, Switzerland
}
\affiliation{Swiss Nanoscience Institute, Klingelbergstrasse 82, 4056 Basel, Switzerland}

\author{Marietta Batzer}
\affiliation{
Department of Physics, University of Basel, Klingelbergstrasse 82, 4056 Basel, Switzerland
}
\affiliation{Swiss Nanoscience Institute, Klingelbergstrasse 82, 4056 Basel, Switzerland}

\author{Marcel.li Grimau}
\affiliation{
Department of Physics, University of Basel, Klingelbergstrasse 82, 4056 Basel, Switzerland
}
\affiliation{Swiss Nanoscience Institute, Klingelbergstrasse 82, 4056 Basel, Switzerland}

\author{Brendan Shields}
\affiliation{
Department of Physics, University of Basel, Klingelbergstrasse 82, 4056 Basel, Switzerland
}
\affiliation{Swiss Nanoscience Institute, Klingelbergstrasse 82, 4056 Basel, Switzerland}

\author{Andrew M.
Edmonds}
\affiliation{
Element Six, Harwell, OX11 0QR, United Kingdom
}

\author{Nicola Palmer}
\affiliation{
Element Six, Harwell, OX11 0QR, United Kingdom
}

\author{Matthew L. Markham}
\affiliation{
Element Six, Harwell, OX11 0QR, United Kingdom
}

\author{Robert J. Cava}
\affiliation{
Department of Chemistry, Princeton University, Princeton, New Jersey 08544, USA
}

\author{Patrick Maletinsky}
\affiliation{
Department of Physics, University of Basel, Klingelbergstrasse 82, 4056 Basel, Switzerland
}
\affiliation{Swiss Nanoscience Institute, Klingelbergstrasse 82, 4056 Basel, Switzerland}

\author{Nathalie P. de Leon}%
 \email{npdeleon@princeton.edu}
\affiliation{%
Department of Electrical and Computer Engineering, Princeton University, Princeton, New Jersey 08544, USA
}%

\date{\today}

\begin{abstract}
Neutral silicon vacancy centers (\siv{}) in diamond are promising candidates for quantum networks because of their long spin coherence times and stable, narrow optical transitions. However, stabilizing \siv{} requires high purity, boron doped diamond, which is not a readily available material. Here, we demonstrate an alternative approach via chemical control of the diamond surface. We use low-damage chemical processing and annealing in a hydrogen environment to realize reversible and highly stable charge state tuning in undoped diamond. The resulting \siv{} centers display optically detected magnetic resonance and bulk-like optical properties. Controlling the charge state tuning via surface termination offers a route for scalable technologies based on \siv{} centers, as well as charge state engineering of other defects.
\end{abstract}

\maketitle


Color centers in diamond are promising platforms for quantum information processing and quantum sensing. As atom-like systems, they can exhibit favorable properties such as long spin coherence times and narrow optical transitions \cite{Gao2015,Awschalom2018,Atature2018}. Aside from their spin and orbital degrees of freedom, these color centers can often exhibit multiple stable charge states. The charge degree of freedom can be used as a powerful resource in spin-to-charge conversion and photoelectric detection \cite{Shields2015,Bourgeois2015}, however uncontrolled charge state conversion can also hinder applications and reduce the fidelity of quantum state manipulation \cite{Yuan2020}. It is therefore of great importance to gain understanding of the mechanisms for charge state conversion and develop methods to stabilize the desired charge state.

Recently, silicon vacancy (SiV) centers in diamond have emerged as a leading platform for quantum network applications \cite{Nguyen2019,Bhaskar2020}. SiV centers are known to exhibit two optically active charge states: negative (\sivm{}) and neutral (\siv{}). \sivm{} centers possess narrow, stable optical transitions, but also exhibit rapid spin decoherence at liquid helium temperatures, requiring sub-Kelvin temperatures for coherent spin manipulation \cite{Jahnke2015,Sukachev2017}. On the other hand, \siv{} centers maintain long spin coherence times at elevated temperatures, and they possess narrow, coherent optical transitions and spin-dependent fluorescence, making them a competitive candidate for quantum networks \cite{Rose2018,Zhang2020}. However, the fabrication of substrates with high conversion efficiency to \siv{} remains challenging. The neutral charge state is not stable in typical high-purity diamonds and requires pinning the Fermi level close to the valence band maximum (VBM) \cite{Gali2013} while maintaining a high purity environment. Scalable fabrication of such substrates remains an outstanding challenge, and high conversion yield of \siv{} has been restricted to a limited number of high purity, boron doped diamonds \cite{Rose2018}.

Here, we demonstrate a novel approach to stabilizing the neutral charge state of near-surface SiV centers in diamond by surface transfer doping. The electronic properties of the diamond surface are known to depend strongly on the surface termination. Specifically, hydrogen-terminated (H-terminated) diamond exhibits a negative electron affinity, pulling the VBM above acceptor levels for surface adsorbates, which leads to a charge transfer process that in turn gives rise to Fermi level pinning near the VBM and band bending \cite{Landstrass1989,Maier2000,Garrido2008}. This surface transfer doping can be used to modulate the charge state of near-surface diamond defects. For example, the negative charge state of nitrogen vacancy (NV) centers was shown to be quenched under H-terminated surfaces \cite{Hauf2011}, and active tuning of the NV charge state was demonstrated under H-terminated surfaces with electrolytic and in-plane gate electrodes \cite{Grotz2012,Karaveli2016,Hauf2014}. In this work, we develop a gentle, non-destructive, and robust approach of modifying the surface termination and demonstrate reversible tuning of the charge state of SiV centers under different surface terminations. We show that the neutral charge state can be generated efficiently under H-termination while the negative charge state is more favorable under oxygen termination (O-termination). We observe bulk-like optical properties and optically detected magnetic resonance (ODMR) of \siv{} centers under H-terminated surfaces, paving the way for scalable fabrication of \siv{} containing substrates in undoped diamond.

A high purity diamond grown by plasma chemical vapor deposition (Element Six ``electronic grade") was used in the experiments. The diamond was polished into a 50 $\mu$m membrane and implanted with $^{28}$Si at 25~keV with total fluence of \mbox{3$\times 10^{11}$ cm$^{-2}$}. The average depth of implanted Si is estimated to be 20~nm using stopping range in matter (SRIM) calculations (Fig.~S1(a) \cite{Supplemental}). The SiV centers were activated using high-temperature vacuum annealing \cite{Evans2016}. We then fabricated parabolic reflectors (PR) with diameters of approximately 300~nm on the diamond membrane to enhance the collection efficiency \cite{Hedrich2020}. The O-terminated surface was prepared with a refluxing mixture of concentrated perchloric, nitric, and sulfuric acids (tri-acid cleaning). The H-terminated surface was prepared by annealing the sample in either pure hydrogen at 750\degree C for 6 hours or in forming gas (5\% H$_2$ in Ar) at 800\degree C for 72 hours. Measurements shown in the main text are based on samples prepared with the latter method. The \siv{} measurements were conducted in a near-infrared confocal microscope at cryogenic temperatures. The \sivm{} measurements were conducted in a visible wavelength confocal microscope at room temperature in ambient conditions. Optical emission was detected with either a spectrometer (Princeton Instruments Acton SP-2300i with Pixis 100 CCD and 300 g/mm grating) or a superconducting nanowire detector (Quantum Opus, optimized for 950 - 1100 nm). For more details of the experiments and sample preparation, see supplemental materials \cite{Supplemental}.

\begin{figure}[h!]
  \centering
  \includegraphics[width = 86mm]{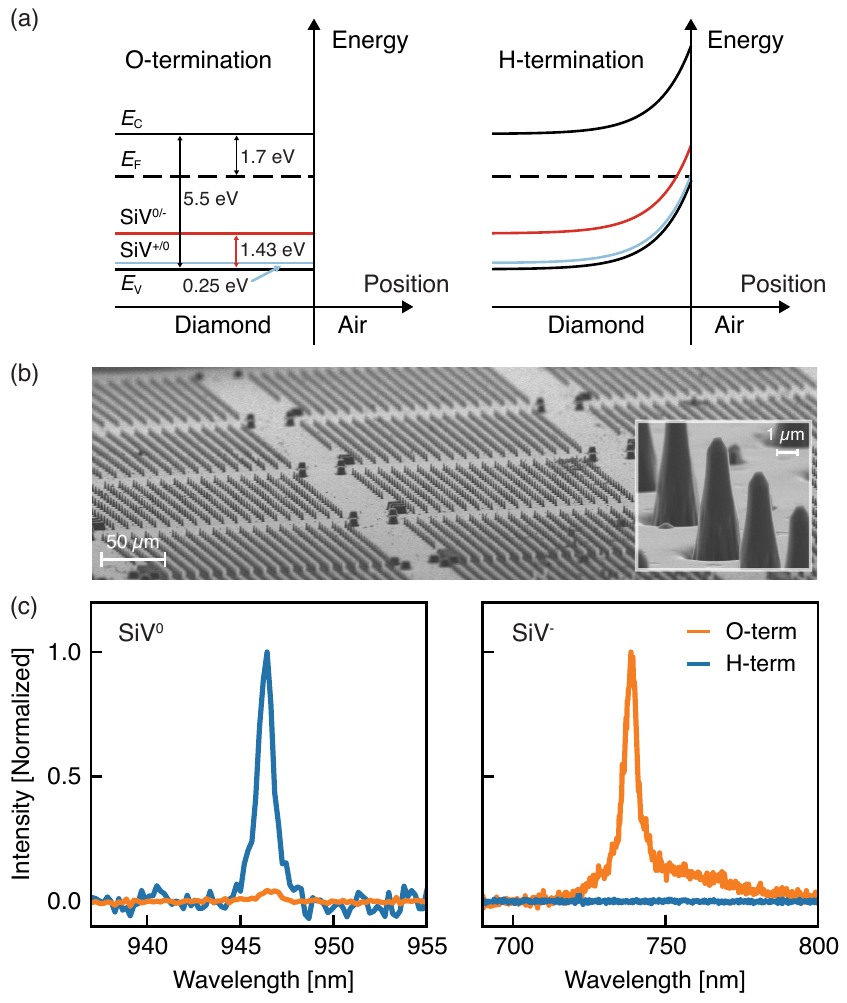}
  \caption{{\bf Charge state tuning of SiV centers via surface termination.}
  (a) Energy band diagram of diamond under different surface terminations. Under O-termination, the Fermi level is dominated by the nitrogen donors (with ionization energy of 1.7~eV) in the bulk. Under H-termination, upwards band bending occurs and the Fermi level near the diamond surface is pinned to the valence band maximum ($E_\textrm{V}$). $E_\textrm{C}$ and $E_\textrm{F}$ represent the positions of the conduction band minimum and Fermi level of diamond. The SiV$^{+/0}$ charge transition level ($E_\textrm{V}$ + 0.25~eV) is shown as the blue curve, and the SiV$^{0/-}$ charge transition level ($E_\textrm{V}$ + 1.43~eV) is shown as the red curve. (b) Scanning electron microscope (SEM) image of the PRs hosting the SiV centers (80$\degree$ tilt). Inset: high resolution SEM of PRs. (c) Characteristic emission spectrum of SiV centers in a representative PR under different terminations. \siv{} spectra were taken using 10~mW 857~nm excitation at 70~K. \sivm{} spectra were taken with 0.1~mW 561 nm excitation at room temperature.}
  \label{fig:Fig1}
\end{figure}

The dominant charge state of SiV centers is determined by the relative position between the Fermi level and SiV charge transition levels (Fig.~\ref{fig:Fig1}(a)). The charge transition levels for SiV centers were calculated to be 0.25~eV above the VBM for SiV$^{+/0}$ and 1.43~eV above the VBM for SiV$^{0/-}$ \cite{Gali2013,Thiering2018}. For an O-terminated surface, the Fermi level near the surface is dominated by nitrogen donors in the bulk (concentration below 5 ppb) that pin the Fermi level to 1.7~eV below the conduction band minimum (CBM) \cite{Farrer1969}, and the negative charge state will be thermodynamically favored. Prior work on stabilizing the neutral charge state relies on boron doping \cite{Rose2018}, which pins the Fermi level at 0.37~eV above the VBM \cite{Chrenko1973}, in between SiV$^{0/-}$ and SiV$^{+/0}$. An alternative approach is to use high purity, undoped diamond with an H-terminated surface. For an H-terminated surface, charge transfer from the valence band to surface adsorbates leads to the accumulation of a two dimensional hole gas at the diamond surface, which pins the Fermi level to the VBM and results in upward band bending, pulling the SiV$^{0/-}$ charge transition point above the Fermi level \cite{Landstrass1989,Maier2000,Garrido2008}. With 5 ppb nitrogen as the dominant donor, the length scale of the band bending can extend beyond 50~nm \cite{Hauf2011}.

We study the charge state behavior of SiV centers in a membrane sample with a two dimensional array of PRs (Fig.~\ref{fig:Fig1}(b)). Focusing on the photoluminescence spectrum of SiV centers in a particular PR, we observe that the emission spectrum changes drastically upon changing the chemical termination of the surface (Fig.~\ref{fig:Fig1}(c)). Compared to the O-terminated surface, after H-termination we observe that the \siv{} emission increases while the \sivm{} emission decreases below background levels. We also observe formation of \siv{} in unetched regions of the membrane under H-termination (Fig.~S6 \cite{Supplemental}). Our results are consistent with prior work on H-terminated nanodiamonds where significant decrease in total \sivm{} fluorescence is observed after H-termination via hydrogen plasma \cite{Rogers2019}; however in prior work there has been no reported observation of \siv{}. The observed emission intensity of \siv{} formed under H-termination in the uneteched region is within a factor of four of the emission intensity of \siv{} prepared in a boron doped diamond with the same implantation dose, indicating that surface control is comparably effective to bulk doping (Fig.~S7(a) \cite{Supplemental}).

\begin{figure}[h!]
  \centering
  \includegraphics[width = 86mm]{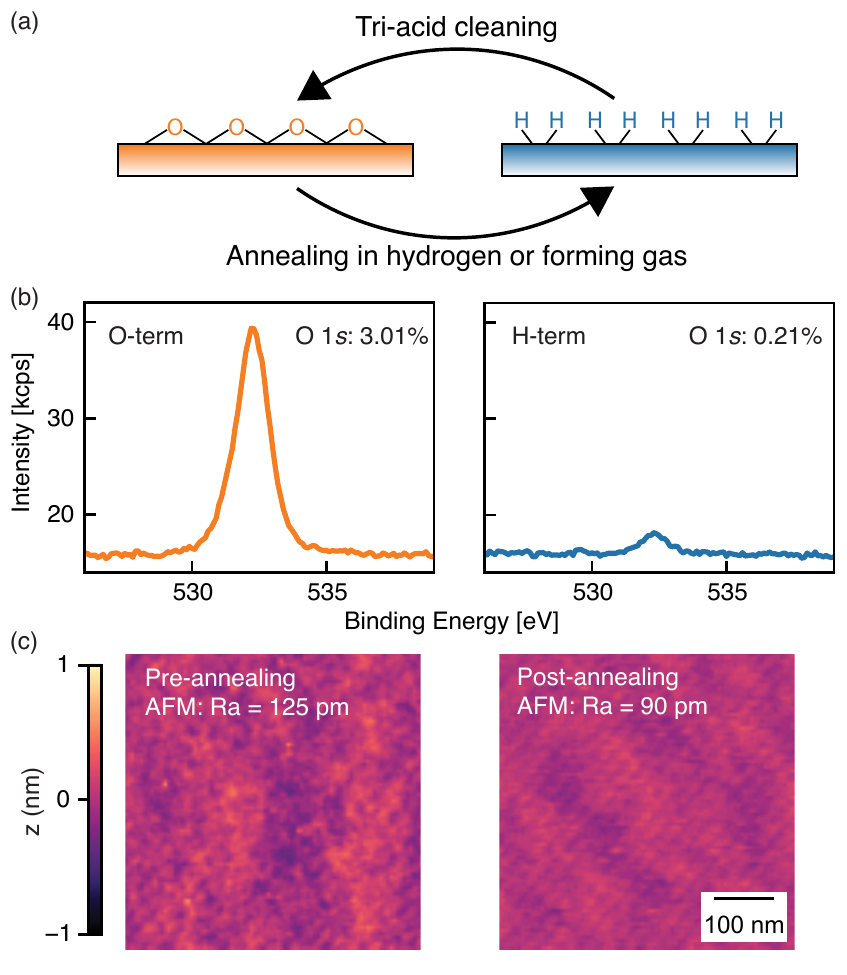}
  \caption{{\bf Characterization of the surface termination process.}
  (a) Schematic illustration of the surface termination process. The O-terminated surface is prepared using tri-acid cleaning. The H-terminated surface is prepared by annealing the sample in forming gas at 800\degree C for 72 hours or in hydrogen at 750\degree C for 6 hours. (b) XPS of the oxygen 1$s$ peak in a test sample after the different processes. Hydrogen cannot be detected using XPS. Therefore, the absence of the oxygen 1$s$ peak and any peaks other than the carbon 1$s$ peak indicates the presence of H-termination. We measure XPS of the sample immediately after the processes to minimize contribution from contamination on the surface. (c) AFM scans of the surface of a test diamond before and after the annealing. The surface morphology is unaffected by the annealing process.}
  \label{fig:Fig2}
\end{figure}

In order to preserve the properties of near-surface color centers, it is important to use a gentle surface termination procedure that avoids subsurface damage and surface roughening. To prepare an H-terminated surface, we first ensure that the diamond surface is contamination free by tri-acid cleaning, and then we anneal the sample either in hydrogen or in forming gas (Fig.~\ref{fig:Fig2}(a)). To remove the H-termination and reset the surface to O-termination, we perform tri-acid cleaning again. X-ray photoelectron spectroscopy (XPS) shows that the oxygen 1$s$ peak intensity decreases after hydrogen termination (Fig.~\ref{fig:Fig2}(b)). From the inelastic mean free path of photoelectrons and the X-ray energy (1487~eV), we estimate the contribution of the signal from a monolayer of atoms on diamond surface to be 7.6\% \cite{Shinotsuka2015, Peace2019}. Based on the oxygen 1$s$ signal intensity observed in XPS, we conclude that the surface has $\sim$40\% of a monolayer of oxygen after tri-acid cleaning while the oxygen coverage is less than $\sim$3\% of a monolayer after hydrogen termination. The sub-monolayer coverage of oxygen after tri-acid cleaning is consistent with previous observations \cite{Peace2019}. This surface termination procedure is reversible and shows consistent results over many rounds of tri-acid cleaning and hydrogen termination (Fig.~S2 \cite{Supplemental}).
Atomic force microscopy (AFM) scans before and after forming gas annealing reveal that the surface remains smooth (Fig.~\ref{fig:Fig2}(c)), making the process compatible with shallow color center applications. This is in stark contrast to the commonly used hydrogen plasma treatment where significant surface roughening, subsurface damage, or hydrogen atom diffusion may occur during the plasma treatment \cite{Koslowski1998,Gaisinskaya2009,Crawford2018,Stacey2012}.
 
 \begin{figure}[h!]
  \centering
  \includegraphics[width = 86mm]{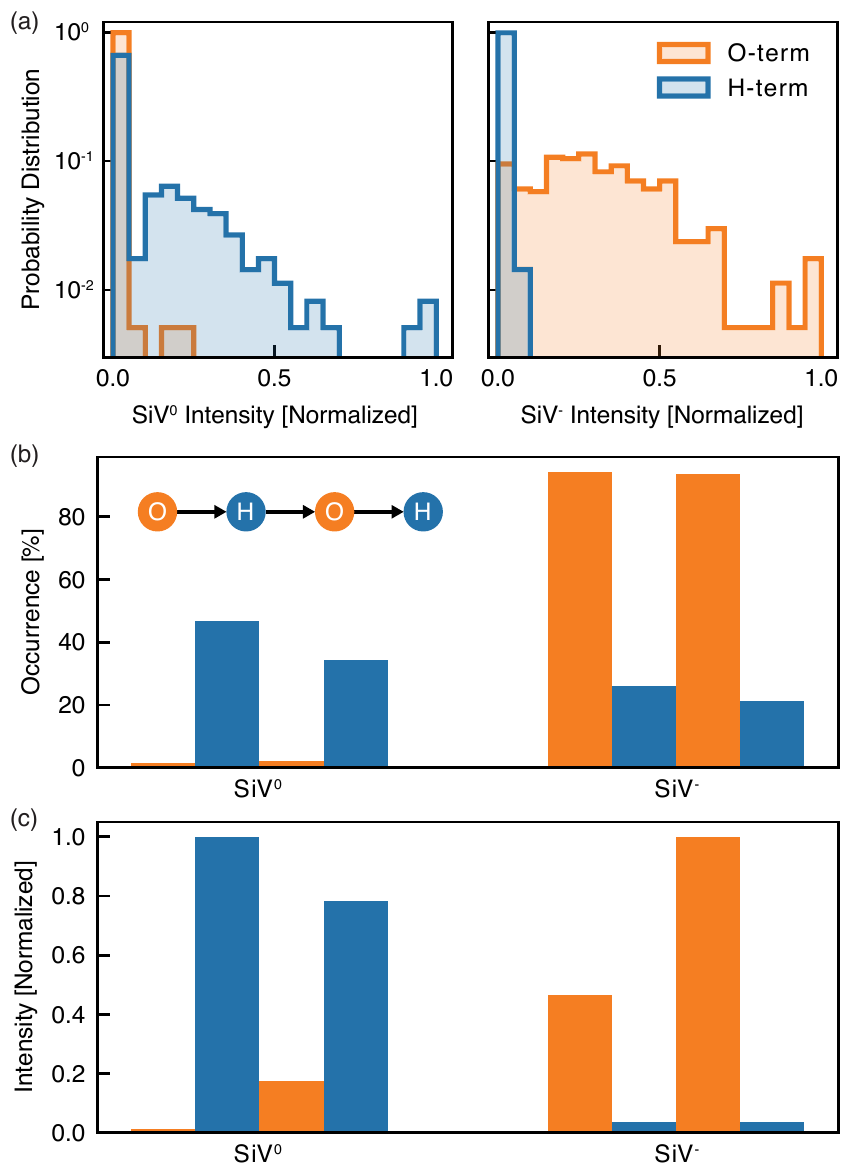}
  \caption{{\bf Statistics of SiV emission under different surface terminations.}
  (a) Left: \siv{} emission intensity distribution under H- and O-termination measured at 10~K with 11~mW 857~nm excitation. Right: \sivm{} emission intensity distribution under H- and O-termination measured in air with 0.1~mW 561~nm excitation. The intensities are extracted from Lorentzian fits of zero-phonon line intensities. Intensity for PRs without observable SiV emission is set to zero. (b) Occurrence (ratio of the number of PRs with SiV emission above the background to total number of PRs) of \siv{} and \sivm{} centers under different surface terminations. Orange denotes O-termination while blue denotes H-termination. Inset: the order of surface termination under study. (c) Average intensity of \siv{} and \sivm{} centers under different surface termination. \siv{} emission is suppressed under O-termination while \sivm{} emission is suppressed under H-termination. The PRs where no emission was identified are excluded from the analysis. \sivm{} centers were measured using 1.88 mW 561 nm excitation for the first O-terminated surface while the power was kept at 0.1 mW for the rest of the measurements. The intensities are scaled by power.}
  \label{fig:Fig3}
\end{figure}

We quantify the effect of surface termination on SiV centers by measuring spectral statistics across a large number of PRs. Histograms of the \siv{} and \sivm{} emission intensities are dramatically different under the two different surface terminations (Fig.~\ref{fig:Fig3}(a)). Specifically, we observe that \siv{} center emission is suppressed under O-termination while \sivm{} center emission is suppressed under H-termination. We iteratively prepare the surface with O-termination and H-termination and observe reversible tuning between \sivm{} and \siv{}.  We observe that both the occurrence (Fig.~\ref{fig:Fig3}(b)) and intensity (Fig.~\ref{fig:Fig3}(c)) toggle reversibly for \sivm{} and \siv{} centers.  This is in contrast to plasma-based processes, where diffusion of atomic hydrogen can lead to irreversible depletion of NV centers \cite{Stacey2012}. In addition, we probe the stability of the surface against cleaning in piranha solution (a 1:2 mixture of hydrogen peroxide in concentrated sulfuric acid) and long-term air exposure. The spectral statistics of SiV centers are unchanged upon piranha cleaning (Fig.~S8 \cite{Supplemental}). In fact, we note that initial observation of \siv{} was under a surface prepared by annealing in hydrogen followed by more than 1.5 years of air exposure (Fig.~S3 \cite{Supplemental}). The stability of the charge state distribution against piranha cleaning and air exposure demonstrates the exceptional robustness of this approach.

\begin{figure}[h!]
  \centering
  \includegraphics[width = 86mm]{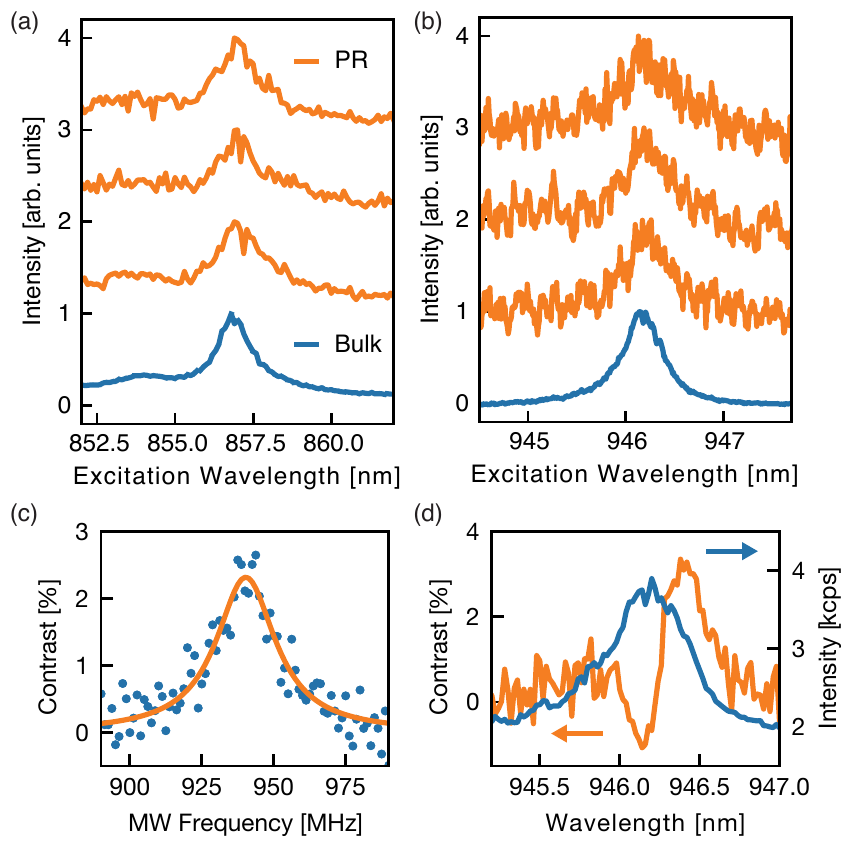}
  \caption{{\bf PLE and ODMR of \siv{} under H-termination.}
  (a) PLE of the bound exciton transition on \siv{} in three H-terminated PRs (orange) and a bulk-doped sample (blue). The excitation power is $\sim$10~mW. (b) PLE of the zero-phonon line transition of \siv{} in three H-terminated PRs (orange) and a bulk-doped sample (blue). The PLE curves are background subtracted. The excitation power is $\sim$0.3~mW. The curves in (a) and (b) are offset from each other for clarity. (c) Continuous-wave ODMR of \siv{} centers in a PR measured by exciting at 946.45 nm at 10~K. The orange curve shows a Lorentzian fit to the data. (d) Wavelength dependence of ODMR contrast (orange curve). Blue curve shows the PLE spectrum for comparison. ODMR was measured after H-termination and three piranha cleaning steps with the excitation power set to $\sim$25 $\mu$W.}
  \label{fig:Fig4}
\end{figure}

Finally, we demonstrate that \siv{} centers prepared under an H-terminated surface show similar properties to samples where \siv{} centers are formed by bulk silicon doping during growth \cite{Rose2018,Zhang2020}. We investigate the optical transitions of \siv{} hosted in PRs prepared under an H-terminated surface using photoluminescence excitation (PLE) spectroscopy. First, we scan a narrow linewidth laser across one of the bound exciton transitions of \siv{} \cite{Zhang2020} while monitoring the emission into the zero-phonon line (ZPL) at 946~nm. A resonance around 857~nm is observed, consistent with the spectrum from a bulk doped sample, as shown in Fig.~\ref{fig:Fig4}(a). Then, we probe the ZPL transition by scanning a narrow linewidth laser across the \siv{} ZPL while emission is measured at wavelengths longer than 960~nm. We observe a resonance at 946.2~nm, consistent with bulk PLE measurements (Fig.~\ref{fig:Fig4}(b)).

We also observe ODMR in \siv{} centers prepared under an H-terminated surface via excitation of the ZPL transition at 946.45~nm. As the microwave (MW) frequency is swept across the zero-field splitting of \siv{}, we observe a resonance peak at 940 MHz (Fig.~\ref{fig:Fig4}(c)). In addition, we observe that the ODMR contrast at the ZPL is strongly wavelength dependent (Fig.~\ref{fig:Fig4}(d)), suggesting that even though individual transitions are not resolvable in PLE, they are spectrally separated enough to allow spin-dependent fluorescence. The observation of ODMR in \siv{} centers via the ZPL transition is complementary to the recently demonstrated ODMR via the bound exciton transitions \cite{Zhang2020}. We note that the observed ODMR frequency of 940 MHz is slightly lower than the previously reported zero-field splitting \cite{Edmonds2008}, and that in this sample the ODMR frequency shifts with higher optical excitation power and MW power (Fig.~S10 and Fig.~S11 \cite{Supplemental}). The origin of the shift is currently under investigation, but we note that the sign of the shift is inconsistent with heating from the microwaves or laser \cite{Edmonds2008}. 

In conclusion, we have demonstrated that chemical control of the diamond surface can be used to tune the charge state of SiV centers to stabilize the neutral charge state in undoped diamond. The gentle surface termination procedure we developed here allows for non-destructive, reversible and long-lived control of the diamond surface. Near-surface \siv{} centers prepared using our approach preserve bulk-like optical properties and allow for optically detected magnetic resonance.

Our approach provides an alternative route to controlling color centers in diamond without careful control over the doping or defect states in the bulk. In addition, the present work focuses on static stabilization of a particular charge state. Utilizing the surface termination to control the charge state may also be compatible with dynamic electric field tuning of the Fermi level, as has been demonstrated for NV centers with electrolytic and in-plane gate electrodes \cite{Grotz2012,Karaveli2016,Hauf2014}. Such an approach could be widely applicable to other color centers in diamond. For example, Fermi level tuning via surface chemical control could help stabilize and identify other neutral group IV vacancy centers, or access additional charge states of SiV (SiV${^+}$ and SiV$^{2-}$), whose spectroscopic signatures have been elusive \cite{Krivobok2020,Luhmann2020}, possibly due to challenges in Fermi level engineering. Another avenue of exploration is to control the charge transfer process by using explicit electron acceptor materials on H-terminated diamond, for example molecular species (NO$_2$, C$_{60}$ and O$_3$ \cite{SATO201347,Kubovic2010,Strobel2004}) or solid encapsulation materials such as transition metal oxides \cite{Russell2013}.

\begin{acknowledgments}
We gratefully acknowledge Z.~Yuan for use of the visible wavelength confocal microscope and S.~Mukherjee for help with surface characterization. Spectroscopy of SiV was supported by National Science Foundation through the Princeton Center for Complex Materials, a Materials Research Science and Engineering Center (Grant No. DMR-1420541) and the Air Force Office of Scientific Research under Grant No. FA9550-17-0158. Surface processing to control the charge state was supported by the U.S. Department of Energy, Office of Science, National Quantum Information Science Research Centers, Co-design Center for Quantum Advantage (C2QA) under contract number DE-SC0012704, the Swiss Nanoscience Institute and the quantERA grant SensExtreme. L.V.H.R. acknowledges support from the Department of Defense through the National Defense Science and Engineering Graduate Fellowship Program. 
\end{acknowledgments}




\bibliography{ChemicalControl}

\providecommand{\noopsort}[1]{}\providecommand{\singleletter}[1]{#1}%
\begin{thebibliography}{40}%
\makeatletter
\providecommand \@ifxundefined [1]{%
 \@ifx{#1\undefined}
}%
\providecommand \@ifnum [1]{%
 \ifnum #1\expandafter \@firstoftwo
 \else \expandafter \@secondoftwo
 \fi
}%
\providecommand \@ifx [1]{%
 \ifx #1\expandafter \@firstoftwo
 \else \expandafter \@secondoftwo
 \fi
}%
\providecommand \natexlab [1]{#1}%
\providecommand \enquote  [1]{``#1''}%
\providecommand \bibnamefont  [1]{#1}%
\providecommand \bibfnamefont [1]{#1}%
\providecommand \citenamefont [1]{#1}%
\providecommand \href@noop [0]{\@secondoftwo}%
\providecommand \href [0]{\begingroup \@sanitize@url \@href}%
\providecommand \@href[1]{\@@startlink{#1}\@@href}%
\providecommand \@@href[1]{\endgroup#1\@@endlink}%
\providecommand \@sanitize@url [0]{\catcode `\\12\catcode `\$12\catcode
  `\&12\catcode `\#12\catcode `\^12\catcode `\_12\catcode `\%12\relax}%
\providecommand \@@startlink[1]{}%
\providecommand \@@endlink[0]{}%
\providecommand \url  [0]{\begingroup\@sanitize@url \@url }%
\providecommand \@url [1]{\endgroup\@href {#1}{\urlprefix }}%
\providecommand \urlprefix  [0]{URL }%
\providecommand \Eprint [0]{\href }%
\providecommand \doibase [0]{https://doi.org/}%
\providecommand \selectlanguage [0]{\@gobble}%
\providecommand \bibinfo  [0]{\@secondoftwo}%
\providecommand \bibfield  [0]{\@secondoftwo}%
\providecommand \translation [1]{[#1]}%
\providecommand \BibitemOpen [0]{}%
\providecommand \bibitemStop [0]{}%
\providecommand \bibitemNoStop [0]{.\EOS\space}%
\providecommand \EOS [0]{\spacefactor3000\relax}%
\providecommand \BibitemShut  [1]{\csname bibitem#1\endcsname}%
\let\auto@bib@innerbib\@empty
\bibitem [{\citenamefont {Gao}\ \emph {et~al.}(2015)\citenamefont {Gao},
  \citenamefont {Imamoglu}, \citenamefont {Bernien},\ and\ \citenamefont
  {Hanson}}]{Gao2015}%
  \BibitemOpen
  \bibfield  {author} {\bibinfo {author} {\bibfnamefont {W.~B.}\ \bibnamefont
  {Gao}}, \bibinfo {author} {\bibfnamefont {A.}~\bibnamefont {Imamoglu}},
  \bibinfo {author} {\bibfnamefont {H.}~\bibnamefont {Bernien}},\ and\ \bibinfo
  {author} {\bibfnamefont {R.}~\bibnamefont {Hanson}},\ }\bibfield  {title}
  {\bibinfo {title} {Coherent manipulation, measurement and entanglement of
  individual solid-state spins using optical fields},\ }\href
  {https://doi.org/10.1038/nphoton.2015.58} {\bibfield  {journal} {\bibinfo
  {journal} {Nature Photonics}\ }\textbf {\bibinfo {volume} {9}},\ \bibinfo
  {pages} {363} (\bibinfo {year} {2015})}\BibitemShut {NoStop}%
\bibitem [{\citenamefont {Awschalom}\ \emph {et~al.}(2018)\citenamefont
  {Awschalom}, \citenamefont {Hanson}, \citenamefont {Wrachtrup},\ and\
  \citenamefont {Zhou}}]{Awschalom2018}%
  \BibitemOpen
  \bibfield  {author} {\bibinfo {author} {\bibfnamefont {D.~D.}\ \bibnamefont
  {Awschalom}}, \bibinfo {author} {\bibfnamefont {R.}~\bibnamefont {Hanson}},
  \bibinfo {author} {\bibfnamefont {J.}~\bibnamefont {Wrachtrup}},\ and\
  \bibinfo {author} {\bibfnamefont {B.~B.}\ \bibnamefont {Zhou}},\ }\bibfield
  {title} {\bibinfo {title} {Quantum technologies with optically interfaced
  solid-state spins},\ }\href {https://doi.org/10.1038/s41566-018-0232-2}
  {\bibfield  {journal} {\bibinfo  {journal} {Nature Photonics}\ }\textbf
  {\bibinfo {volume} {12}},\ \bibinfo {pages} {516} (\bibinfo {year}
  {2018})}\BibitemShut {NoStop}%
\bibitem [{\citenamefont {Atat{\"u}re}\ \emph {et~al.}(2018)\citenamefont
  {Atat{\"u}re}, \citenamefont {Englund}, \citenamefont {Vamivakas},
  \citenamefont {Lee},\ and\ \citenamefont {Wrachtrup}}]{Atature2018}%
  \BibitemOpen
  \bibfield  {author} {\bibinfo {author} {\bibfnamefont {M.}~\bibnamefont
  {Atat{\"u}re}}, \bibinfo {author} {\bibfnamefont {D.}~\bibnamefont
  {Englund}}, \bibinfo {author} {\bibfnamefont {N.}~\bibnamefont {Vamivakas}},
  \bibinfo {author} {\bibfnamefont {S.-Y.}\ \bibnamefont {Lee}},\ and\ \bibinfo
  {author} {\bibfnamefont {J.}~\bibnamefont {Wrachtrup}},\ }\bibfield  {title}
  {\bibinfo {title} {Material platforms for spin-based photonic quantum
  technologies},\ }\href {https://doi.org/10.1038/s41578-018-0008-9} {\bibfield
   {journal} {\bibinfo  {journal} {Nature Reviews Materials}\ }\textbf
  {\bibinfo {volume} {3}},\ \bibinfo {pages} {38} (\bibinfo {year}
  {2018})}\BibitemShut {NoStop}%
\bibitem [{\citenamefont {Shields}\ \emph {et~al.}(2015)\citenamefont
  {Shields}, \citenamefont {Unterreithmeier}, \citenamefont {de~Leon},
  \citenamefont {Park},\ and\ \citenamefont {Lukin}}]{Shields2015}%
  \BibitemOpen
  \bibfield  {author} {\bibinfo {author} {\bibfnamefont {B.~J.}\ \bibnamefont
  {Shields}}, \bibinfo {author} {\bibfnamefont {Q.~P.}\ \bibnamefont
  {Unterreithmeier}}, \bibinfo {author} {\bibfnamefont {N.~P.}\ \bibnamefont
  {de~Leon}}, \bibinfo {author} {\bibfnamefont {H.}~\bibnamefont {Park}},\ and\
  \bibinfo {author} {\bibfnamefont {M.~D.}\ \bibnamefont {Lukin}},\ }\bibfield
  {title} {\bibinfo {title} {Efficient readout of a single spin state in
  diamond via spin-to-charge conversion},\ }\href
  {https://doi.org/10.1103/PhysRevLett.114.136402} {\bibfield  {journal}
  {\bibinfo  {journal} {Phys. Rev. Lett.}\ }\textbf {\bibinfo {volume} {114}},\
  \bibinfo {pages} {136402} (\bibinfo {year} {2015})}\BibitemShut {NoStop}%
\bibitem [{\citenamefont {Bourgeois}\ \emph {et~al.}(2015)\citenamefont
  {Bourgeois}, \citenamefont {Jarmola}, \citenamefont {Siyushev}, \citenamefont
  {Gulka}, \citenamefont {Hruby}, \citenamefont {Jelezko}, \citenamefont
  {Budker},\ and\ \citenamefont {Nesladek}}]{Bourgeois2015}%
  \BibitemOpen
  \bibfield  {author} {\bibinfo {author} {\bibfnamefont {E.}~\bibnamefont
  {Bourgeois}}, \bibinfo {author} {\bibfnamefont {A.}~\bibnamefont {Jarmola}},
  \bibinfo {author} {\bibfnamefont {P.}~\bibnamefont {Siyushev}}, \bibinfo
  {author} {\bibfnamefont {M.}~\bibnamefont {Gulka}}, \bibinfo {author}
  {\bibfnamefont {J.}~\bibnamefont {Hruby}}, \bibinfo {author} {\bibfnamefont
  {F.}~\bibnamefont {Jelezko}}, \bibinfo {author} {\bibfnamefont
  {D.}~\bibnamefont {Budker}},\ and\ \bibinfo {author} {\bibfnamefont
  {M.}~\bibnamefont {Nesladek}},\ }\bibfield  {title} {\bibinfo {title}
  {{Photoelectric detection of electron spin resonance of nitrogen-vacancy
  centres in diamond}},\ }\href {https://doi.org/10.1038/ncomms9577} {\bibfield
   {journal} {\bibinfo  {journal} {Nature Communications}\ }\textbf {\bibinfo
  {volume} {6}},\ \bibinfo {pages} {8577} (\bibinfo {year} {2015})}\BibitemShut
  {NoStop}%
\bibitem [{\citenamefont {Yuan}\ \emph {et~al.}(2020)\citenamefont {Yuan},
  \citenamefont {Fitzpatrick}, \citenamefont {Rodgers}, \citenamefont
  {Sangtawesin}, \citenamefont {Srinivasan},\ and\ \citenamefont
  {de~Leon}}]{Yuan2020}%
  \BibitemOpen
  \bibfield  {author} {\bibinfo {author} {\bibfnamefont {Z.}~\bibnamefont
  {Yuan}}, \bibinfo {author} {\bibfnamefont {M.}~\bibnamefont {Fitzpatrick}},
  \bibinfo {author} {\bibfnamefont {L.~V.~H.}\ \bibnamefont {Rodgers}},
  \bibinfo {author} {\bibfnamefont {S.}~\bibnamefont {Sangtawesin}}, \bibinfo
  {author} {\bibfnamefont {S.}~\bibnamefont {Srinivasan}},\ and\ \bibinfo
  {author} {\bibfnamefont {N.~P.}\ \bibnamefont {de~Leon}},\ }\bibfield
  {title} {\bibinfo {title} {Charge state dynamics and optically detected
  electron spin resonance contrast of shallow nitrogen-vacancy centers in
  diamond},\ }\href {https://doi.org/10.1103/PhysRevResearch.2.033263}
  {\bibfield  {journal} {\bibinfo  {journal} {Phys. Rev. Research}\ }\textbf
  {\bibinfo {volume} {2}},\ \bibinfo {pages} {033263} (\bibinfo {year}
  {2020})}\BibitemShut {NoStop}%
\bibitem [{\citenamefont {Nguyen}\ \emph {et~al.}(2019)\citenamefont {Nguyen},
  \citenamefont {Sukachev}, \citenamefont {Bhaskar}, \citenamefont {Machielse},
  \citenamefont {Levonian}, \citenamefont {Knall}, \citenamefont {Stroganov},
  \citenamefont {Riedinger}, \citenamefont {Park}, \citenamefont
  {Lon\ifmmode~\check{c}\else \v{c}\fi{}ar},\ and\ \citenamefont
  {Lukin}}]{Nguyen2019}%
  \BibitemOpen
  \bibfield  {author} {\bibinfo {author} {\bibfnamefont {C.~T.}\ \bibnamefont
  {Nguyen}}, \bibinfo {author} {\bibfnamefont {D.~D.}\ \bibnamefont
  {Sukachev}}, \bibinfo {author} {\bibfnamefont {M.~K.}\ \bibnamefont
  {Bhaskar}}, \bibinfo {author} {\bibfnamefont {B.}~\bibnamefont {Machielse}},
  \bibinfo {author} {\bibfnamefont {D.~S.}\ \bibnamefont {Levonian}}, \bibinfo
  {author} {\bibfnamefont {E.~N.}\ \bibnamefont {Knall}}, \bibinfo {author}
  {\bibfnamefont {P.}~\bibnamefont {Stroganov}}, \bibinfo {author}
  {\bibfnamefont {R.}~\bibnamefont {Riedinger}}, \bibinfo {author}
  {\bibfnamefont {H.}~\bibnamefont {Park}}, \bibinfo {author} {\bibfnamefont
  {M.}~\bibnamefont {Lon\ifmmode~\check{c}\else \v{c}\fi{}ar}},\ and\ \bibinfo
  {author} {\bibfnamefont {M.~D.}\ \bibnamefont {Lukin}},\ }\bibfield  {title}
  {\bibinfo {title} {Quantum network nodes based on diamond qubits with an
  efficient nanophotonic interface},\ }\href
  {https://doi.org/10.1103/PhysRevLett.123.183602} {\bibfield  {journal}
  {\bibinfo  {journal} {Phys. Rev. Lett.}\ }\textbf {\bibinfo {volume} {123}},\
  \bibinfo {pages} {183602} (\bibinfo {year} {2019})}\BibitemShut {NoStop}%
\bibitem [{\citenamefont {Bhaskar}\ \emph {et~al.}(2020)\citenamefont
  {Bhaskar}, \citenamefont {Riedinger}, \citenamefont {Machielse},
  \citenamefont {Levonian}, \citenamefont {Nguyen}, \citenamefont {Knall},
  \citenamefont {Park}, \citenamefont {Englund}, \citenamefont {Lon{\v{c}}ar},
  \citenamefont {Sukachev},\ and\ \citenamefont {Lukin}}]{Bhaskar2020}%
  \BibitemOpen
  \bibfield  {author} {\bibinfo {author} {\bibfnamefont {M.~K.}\ \bibnamefont
  {Bhaskar}}, \bibinfo {author} {\bibfnamefont {R.}~\bibnamefont {Riedinger}},
  \bibinfo {author} {\bibfnamefont {B.}~\bibnamefont {Machielse}}, \bibinfo
  {author} {\bibfnamefont {D.~S.}\ \bibnamefont {Levonian}}, \bibinfo {author}
  {\bibfnamefont {C.~T.}\ \bibnamefont {Nguyen}}, \bibinfo {author}
  {\bibfnamefont {E.~N.}\ \bibnamefont {Knall}}, \bibinfo {author}
  {\bibfnamefont {H.}~\bibnamefont {Park}}, \bibinfo {author} {\bibfnamefont
  {D.}~\bibnamefont {Englund}}, \bibinfo {author} {\bibfnamefont
  {M.}~\bibnamefont {Lon{\v{c}}ar}}, \bibinfo {author} {\bibfnamefont {D.~D.}\
  \bibnamefont {Sukachev}},\ and\ \bibinfo {author} {\bibfnamefont {M.~D.}\
  \bibnamefont {Lukin}},\ }\bibfield  {title} {\bibinfo {title} {{Experimental
  demonstration of memory-enhanced quantum communication}},\ }\href
  {https://doi.org/10.1038/s41586-020-2103-5} {\bibfield  {journal} {\bibinfo
  {journal} {Nature}\ }\textbf {\bibinfo {volume} {580}},\ \bibinfo {pages}
  {60} (\bibinfo {year} {2020})}\BibitemShut {NoStop}%
\bibitem [{\citenamefont {Jahnke}\ \emph {et~al.}(2015)\citenamefont {Jahnke},
  \citenamefont {Sipahigil}, \citenamefont {Binder}, \citenamefont {Doherty},
  \citenamefont {Metsch}, \citenamefont {Rogers}, \citenamefont {Manson},
  \citenamefont {Lukin},\ and\ \citenamefont {Jelezko}}]{Jahnke2015}%
  \BibitemOpen
  \bibfield  {author} {\bibinfo {author} {\bibfnamefont {K.~D.}\ \bibnamefont
  {Jahnke}}, \bibinfo {author} {\bibfnamefont {A.}~\bibnamefont {Sipahigil}},
  \bibinfo {author} {\bibfnamefont {J.~M.}\ \bibnamefont {Binder}}, \bibinfo
  {author} {\bibfnamefont {M.~W.}\ \bibnamefont {Doherty}}, \bibinfo {author}
  {\bibfnamefont {M.}~\bibnamefont {Metsch}}, \bibinfo {author} {\bibfnamefont
  {L.~J.}\ \bibnamefont {Rogers}}, \bibinfo {author} {\bibfnamefont {N.~B.}\
  \bibnamefont {Manson}}, \bibinfo {author} {\bibfnamefont {M.~D.}\
  \bibnamefont {Lukin}},\ and\ \bibinfo {author} {\bibfnamefont
  {F.}~\bibnamefont {Jelezko}},\ }\bibfield  {title} {\bibinfo {title}
  {Electron{\textendash}phonon processes of the silicon-vacancy centre in
  diamond},\ }\href {https://doi.org/10.1088/1367-2630/17/4/043011} {\bibfield
  {journal} {\bibinfo  {journal} {New Journal of Physics}\ }\textbf {\bibinfo
  {volume} {17}},\ \bibinfo {pages} {043011} (\bibinfo {year}
  {2015})}\BibitemShut {NoStop}%
\bibitem [{\citenamefont {Sukachev}\ \emph {et~al.}(2017)\citenamefont
  {Sukachev}, \citenamefont {Sipahigil}, \citenamefont {Nguyen}, \citenamefont
  {Bhaskar}, \citenamefont {Evans}, \citenamefont {Jelezko},\ and\
  \citenamefont {Lukin}}]{Sukachev2017}%
  \BibitemOpen
  \bibfield  {author} {\bibinfo {author} {\bibfnamefont {D.~D.}\ \bibnamefont
  {Sukachev}}, \bibinfo {author} {\bibfnamefont {A.}~\bibnamefont {Sipahigil}},
  \bibinfo {author} {\bibfnamefont {C.~T.}\ \bibnamefont {Nguyen}}, \bibinfo
  {author} {\bibfnamefont {M.~K.}\ \bibnamefont {Bhaskar}}, \bibinfo {author}
  {\bibfnamefont {R.~E.}\ \bibnamefont {Evans}}, \bibinfo {author}
  {\bibfnamefont {F.}~\bibnamefont {Jelezko}},\ and\ \bibinfo {author}
  {\bibfnamefont {M.~D.}\ \bibnamefont {Lukin}},\ }\bibfield  {title} {\bibinfo
  {title} {Silicon-vacancy spin qubit in diamond: A quantum memory exceeding 10
  ms with single-shot state readout},\ }\href
  {https://doi.org/10.1103/PhysRevLett.119.223602} {\bibfield  {journal}
  {\bibinfo  {journal} {Phys. Rev. Lett.}\ }\textbf {\bibinfo {volume} {119}},\
  \bibinfo {pages} {223602} (\bibinfo {year} {2017})}\BibitemShut {NoStop}%
\bibitem [{\citenamefont {Rose}\ \emph {et~al.}(2018)\citenamefont {Rose},
  \citenamefont {Huang}, \citenamefont {Zhang}, \citenamefont {Stevenson},
  \citenamefont {Tyryshkin}, \citenamefont {Sangtawesin}, \citenamefont
  {Srinivasan}, \citenamefont {Loudin}, \citenamefont {Markham}, \citenamefont
  {Edmonds}, \citenamefont {Twitchen}, \citenamefont {Lyon},\ and\
  \citenamefont {de~Leon}}]{Rose2018}%
  \BibitemOpen
  \bibfield  {author} {\bibinfo {author} {\bibfnamefont {B.~C.}\ \bibnamefont
  {Rose}}, \bibinfo {author} {\bibfnamefont {D.}~\bibnamefont {Huang}},
  \bibinfo {author} {\bibfnamefont {Z.-H.}\ \bibnamefont {Zhang}}, \bibinfo
  {author} {\bibfnamefont {P.}~\bibnamefont {Stevenson}}, \bibinfo {author}
  {\bibfnamefont {A.~M.}\ \bibnamefont {Tyryshkin}}, \bibinfo {author}
  {\bibfnamefont {S.}~\bibnamefont {Sangtawesin}}, \bibinfo {author}
  {\bibfnamefont {S.}~\bibnamefont {Srinivasan}}, \bibinfo {author}
  {\bibfnamefont {L.}~\bibnamefont {Loudin}}, \bibinfo {author} {\bibfnamefont
  {M.~L.}\ \bibnamefont {Markham}}, \bibinfo {author} {\bibfnamefont {A.~M.}\
  \bibnamefont {Edmonds}}, \bibinfo {author} {\bibfnamefont {D.~J.}\
  \bibnamefont {Twitchen}}, \bibinfo {author} {\bibfnamefont {S.~A.}\
  \bibnamefont {Lyon}},\ and\ \bibinfo {author} {\bibfnamefont {N.~P.}\
  \bibnamefont {de~Leon}},\ }\bibfield  {title} {\bibinfo {title} {Observation
  of an environmentally insensitive solid-state spin defect in diamond},\
  }\href {https://doi.org/10.1126/science.aao0290} {\bibfield  {journal}
  {\bibinfo  {journal} {Science}\ }\textbf {\bibinfo {volume} {361}},\ \bibinfo
  {pages} {60} (\bibinfo {year} {2018})}\BibitemShut {NoStop}%
\bibitem [{\citenamefont {Zhang}\ \emph {et~al.}(2020)\citenamefont {Zhang},
  \citenamefont {Stevenson}, \citenamefont {Thiering}, \citenamefont {Rose},
  \citenamefont {Huang}, \citenamefont {Edmonds}, \citenamefont {Markham},
  \citenamefont {Lyon}, \citenamefont {Gali},\ and\ \citenamefont
  {de~Leon}}]{Zhang2020}%
  \BibitemOpen
  \bibfield  {author} {\bibinfo {author} {\bibfnamefont {Z.-H.}\ \bibnamefont
  {Zhang}}, \bibinfo {author} {\bibfnamefont {P.}~\bibnamefont {Stevenson}},
  \bibinfo {author} {\bibfnamefont {G.}~\bibnamefont {Thiering}}, \bibinfo
  {author} {\bibfnamefont {B.~C.}\ \bibnamefont {Rose}}, \bibinfo {author}
  {\bibfnamefont {D.}~\bibnamefont {Huang}}, \bibinfo {author} {\bibfnamefont
  {A.~M.}\ \bibnamefont {Edmonds}}, \bibinfo {author} {\bibfnamefont {M.~L.}\
  \bibnamefont {Markham}}, \bibinfo {author} {\bibfnamefont {S.~A.}\
  \bibnamefont {Lyon}}, \bibinfo {author} {\bibfnamefont {A.}~\bibnamefont
  {Gali}},\ and\ \bibinfo {author} {\bibfnamefont {N.~P.}\ \bibnamefont
  {de~Leon}},\ }\bibfield  {title} {\bibinfo {title} {Optically detected
  magnetic resonance in neutral silicon vacancy centers in diamond via bound
  exciton states},\ }\href {https://doi.org/10.1103/PhysRevLett.125.237402}
  {\bibfield  {journal} {\bibinfo  {journal} {Phys. Rev. Lett.}\ }\textbf
  {\bibinfo {volume} {125}},\ \bibinfo {pages} {237402} (\bibinfo {year}
  {2020})}\BibitemShut {NoStop}%
\bibitem [{\citenamefont {Gali}\ and\ \citenamefont {Maze}(2013)}]{Gali2013}%
  \BibitemOpen
  \bibfield  {author} {\bibinfo {author} {\bibfnamefont {A.}~\bibnamefont
  {Gali}}\ and\ \bibinfo {author} {\bibfnamefont {J.~R.}\ \bibnamefont
  {Maze}},\ }\bibfield  {title} {\bibinfo {title} {Ab initio study of the split
  silicon-vacancy defect in diamond: Electronic structure and related
  properties},\ }\href {https://doi.org/10.1103/PhysRevB.88.235205} {\bibfield
  {journal} {\bibinfo  {journal} {Phys. Rev. B}\ }\textbf {\bibinfo {volume}
  {88}},\ \bibinfo {pages} {235205} (\bibinfo {year} {2013})}\BibitemShut
  {NoStop}%
\bibitem [{\citenamefont {Landstrass}\ and\ \citenamefont
  {Ravi}(1989)}]{Landstrass1989}%
  \BibitemOpen
  \bibfield  {author} {\bibinfo {author} {\bibfnamefont {M.~I.}\ \bibnamefont
  {Landstrass}}\ and\ \bibinfo {author} {\bibfnamefont {K.~V.}\ \bibnamefont
  {Ravi}},\ }\bibfield  {title} {\bibinfo {title} {Resistivity of chemical
  vapor deposited diamond films},\ }\href {https://doi.org/10.1063/1.101694}
  {\bibfield  {journal} {\bibinfo  {journal} {Applied Physics Letters}\
  }\textbf {\bibinfo {volume} {55}},\ \bibinfo {pages} {975} (\bibinfo {year}
  {1989})}\BibitemShut {NoStop}%
\bibitem [{\citenamefont {Maier}\ \emph {et~al.}(2000)\citenamefont {Maier},
  \citenamefont {Riedel}, \citenamefont {Mantel}, \citenamefont {Ristein},\
  and\ \citenamefont {Ley}}]{Maier2000}%
  \BibitemOpen
  \bibfield  {author} {\bibinfo {author} {\bibfnamefont {F.}~\bibnamefont
  {Maier}}, \bibinfo {author} {\bibfnamefont {M.}~\bibnamefont {Riedel}},
  \bibinfo {author} {\bibfnamefont {B.}~\bibnamefont {Mantel}}, \bibinfo
  {author} {\bibfnamefont {J.}~\bibnamefont {Ristein}},\ and\ \bibinfo {author}
  {\bibfnamefont {L.}~\bibnamefont {Ley}},\ }\bibfield  {title} {\bibinfo
  {title} {{Origin of surface conductivity in diamond}},\ }\href
  {https://doi.org/10.1103/PhysRevLett.85.3472} {\bibfield  {journal} {\bibinfo
   {journal} {Physical Review Letters}\ }\textbf {\bibinfo {volume} {85}},\
  \bibinfo {pages} {3472} (\bibinfo {year} {2000})}\BibitemShut {NoStop}%
\bibitem [{\citenamefont {Garrido}\ \emph {et~al.}(2008)\citenamefont
  {Garrido}, \citenamefont {Nowy}, \citenamefont {H{\"a}rtl},\ and\
  \citenamefont {Stutzmann}}]{Garrido2008}%
  \BibitemOpen
  \bibfield  {author} {\bibinfo {author} {\bibfnamefont {J.~A.}\ \bibnamefont
  {Garrido}}, \bibinfo {author} {\bibfnamefont {S.}~\bibnamefont {Nowy}},
  \bibinfo {author} {\bibfnamefont {A.}~\bibnamefont {H{\"a}rtl}},\ and\
  \bibinfo {author} {\bibfnamefont {M.}~\bibnamefont {Stutzmann}},\ }\bibfield
  {title} {\bibinfo {title} {The diamond/aqueous electrolyte
  interface:{\thinspace} an impedance investigation},\ }\href
  {https://doi.org/10.1021/la703413y} {\bibfield  {journal} {\bibinfo
  {journal} {Langmuir}\ }\textbf {\bibinfo {volume} {24}},\ \bibinfo {pages}
  {3897} (\bibinfo {year} {2008})}\BibitemShut {NoStop}%
\bibitem [{\citenamefont {Hauf}\ \emph {et~al.}(2011)\citenamefont {Hauf},
  \citenamefont {Grotz}, \citenamefont {Naydenov}, \citenamefont {Dankerl},
  \citenamefont {Pezzagna}, \citenamefont {Meijer}, \citenamefont {Jelezko},
  \citenamefont {Wrachtrup}, \citenamefont {Stutzmann}, \citenamefont
  {Reinhard},\ and\ \citenamefont {Garrido}}]{Hauf2011}%
  \BibitemOpen
  \bibfield  {author} {\bibinfo {author} {\bibfnamefont {M.~V.}\ \bibnamefont
  {Hauf}}, \bibinfo {author} {\bibfnamefont {B.}~\bibnamefont {Grotz}},
  \bibinfo {author} {\bibfnamefont {B.}~\bibnamefont {Naydenov}}, \bibinfo
  {author} {\bibfnamefont {M.}~\bibnamefont {Dankerl}}, \bibinfo {author}
  {\bibfnamefont {S.}~\bibnamefont {Pezzagna}}, \bibinfo {author}
  {\bibfnamefont {J.}~\bibnamefont {Meijer}}, \bibinfo {author} {\bibfnamefont
  {F.}~\bibnamefont {Jelezko}}, \bibinfo {author} {\bibfnamefont
  {J.}~\bibnamefont {Wrachtrup}}, \bibinfo {author} {\bibfnamefont
  {M.}~\bibnamefont {Stutzmann}}, \bibinfo {author} {\bibfnamefont
  {F.}~\bibnamefont {Reinhard}},\ and\ \bibinfo {author} {\bibfnamefont
  {J.~A.}\ \bibnamefont {Garrido}},\ }\bibfield  {title} {\bibinfo {title}
  {Chemical control of the charge state of nitrogen-vacancy centers in
  diamond},\ }\href {https://doi.org/10.1103/PhysRevB.83.081304} {\bibfield
  {journal} {\bibinfo  {journal} {Phys. Rev. B}\ }\textbf {\bibinfo {volume}
  {83}},\ \bibinfo {pages} {081304(R)} (\bibinfo {year} {2011})}\BibitemShut
  {NoStop}%
\bibitem [{\citenamefont {Grotz}\ \emph {et~al.}(2012)\citenamefont {Grotz},
  \citenamefont {Hauf}, \citenamefont {Dankerl}, \citenamefont {Naydenov},
  \citenamefont {Pezzagna}, \citenamefont {Meijer}, \citenamefont {Jelezko},
  \citenamefont {Wrachtrup}, \citenamefont {Stutzmann}, \citenamefont
  {Reinhard},\ and\ \citenamefont {Garrido}}]{Grotz2012}%
  \BibitemOpen
  \bibfield  {author} {\bibinfo {author} {\bibfnamefont {B.}~\bibnamefont
  {Grotz}}, \bibinfo {author} {\bibfnamefont {M.~V.}\ \bibnamefont {Hauf}},
  \bibinfo {author} {\bibfnamefont {M.}~\bibnamefont {Dankerl}}, \bibinfo
  {author} {\bibfnamefont {B.}~\bibnamefont {Naydenov}}, \bibinfo {author}
  {\bibfnamefont {S.}~\bibnamefont {Pezzagna}}, \bibinfo {author}
  {\bibfnamefont {J.}~\bibnamefont {Meijer}}, \bibinfo {author} {\bibfnamefont
  {F.}~\bibnamefont {Jelezko}}, \bibinfo {author} {\bibfnamefont
  {J.}~\bibnamefont {Wrachtrup}}, \bibinfo {author} {\bibfnamefont
  {M.}~\bibnamefont {Stutzmann}}, \bibinfo {author} {\bibfnamefont
  {F.}~\bibnamefont {Reinhard}},\ and\ \bibinfo {author} {\bibfnamefont
  {J.~A.}\ \bibnamefont {Garrido}},\ }\bibfield  {title} {\bibinfo {title}
  {{Charge state manipulation of qubits in diamond}},\ }\href
  {https://doi.org/10.1038/ncomms1729} {\bibfield  {journal} {\bibinfo
  {journal} {Nature Communications}\ }\textbf {\bibinfo {volume} {3}},\
  \bibinfo {pages} {729} (\bibinfo {year} {2012})}\BibitemShut {NoStop}%
\bibitem [{\citenamefont {Karaveli}\ \emph {et~al.}(2016)\citenamefont
  {Karaveli}, \citenamefont {Gaathon}, \citenamefont {Wolcott}, \citenamefont
  {Sakakibara}, \citenamefont {Shemesh}, \citenamefont {Peterka}, \citenamefont
  {Boyden}, \citenamefont {Owen}, \citenamefont {Yuste},\ and\ \citenamefont
  {Englund}}]{Karaveli2016}%
  \BibitemOpen
  \bibfield  {author} {\bibinfo {author} {\bibfnamefont {S.}~\bibnamefont
  {Karaveli}}, \bibinfo {author} {\bibfnamefont {O.}~\bibnamefont {Gaathon}},
  \bibinfo {author} {\bibfnamefont {A.}~\bibnamefont {Wolcott}}, \bibinfo
  {author} {\bibfnamefont {R.}~\bibnamefont {Sakakibara}}, \bibinfo {author}
  {\bibfnamefont {O.~A.}\ \bibnamefont {Shemesh}}, \bibinfo {author}
  {\bibfnamefont {D.~S.}\ \bibnamefont {Peterka}}, \bibinfo {author}
  {\bibfnamefont {E.~S.}\ \bibnamefont {Boyden}}, \bibinfo {author}
  {\bibfnamefont {J.~S.}\ \bibnamefont {Owen}}, \bibinfo {author}
  {\bibfnamefont {R.}~\bibnamefont {Yuste}},\ and\ \bibinfo {author}
  {\bibfnamefont {D.}~\bibnamefont {Englund}},\ }\bibfield  {title} {\bibinfo
  {title} {Modulation of nitrogen vacancy charge state and fluorescence in
  nanodiamonds using electrochemical potential},\ }\href
  {https://doi.org/10.1073/pnas.1504451113} {\bibfield  {journal} {\bibinfo
  {journal} {Proceedings of the National Academy of Sciences}\ }\textbf
  {\bibinfo {volume} {113}},\ \bibinfo {pages} {3938} (\bibinfo {year}
  {2016})}\BibitemShut {NoStop}%
\bibitem [{\citenamefont {Hauf}\ \emph {et~al.}(2014)\citenamefont {Hauf},
  \citenamefont {Simon}, \citenamefont {Aslam}, \citenamefont {Pfender},
  \citenamefont {Neumann}, \citenamefont {Pezzagna}, \citenamefont {Meijer},
  \citenamefont {Wrachtrup}, \citenamefont {Stutzmann}, \citenamefont
  {Reinhard},\ and\ \citenamefont {Garrido}}]{Hauf2014}%
  \BibitemOpen
  \bibfield  {author} {\bibinfo {author} {\bibfnamefont {M.~V.}\ \bibnamefont
  {Hauf}}, \bibinfo {author} {\bibfnamefont {P.}~\bibnamefont {Simon}},
  \bibinfo {author} {\bibfnamefont {N.}~\bibnamefont {Aslam}}, \bibinfo
  {author} {\bibfnamefont {M.}~\bibnamefont {Pfender}}, \bibinfo {author}
  {\bibfnamefont {P.}~\bibnamefont {Neumann}}, \bibinfo {author} {\bibfnamefont
  {S.}~\bibnamefont {Pezzagna}}, \bibinfo {author} {\bibfnamefont
  {J.}~\bibnamefont {Meijer}}, \bibinfo {author} {\bibfnamefont
  {J.}~\bibnamefont {Wrachtrup}}, \bibinfo {author} {\bibfnamefont
  {M.}~\bibnamefont {Stutzmann}}, \bibinfo {author} {\bibfnamefont
  {F.}~\bibnamefont {Reinhard}},\ and\ \bibinfo {author} {\bibfnamefont
  {J.~A.}\ \bibnamefont {Garrido}},\ }\bibfield  {title} {\bibinfo {title}
  {{Addressing Single Nitrogen-Vacancy Centers in Diamond with Transparent
  in-Plane Gate Structures}},\ }\href {https://doi.org/10.1021/nl4047619}
  {\bibfield  {journal} {\bibinfo  {journal} {Nano Letters}\ }\textbf {\bibinfo
  {volume} {14}},\ \bibinfo {pages} {2359} (\bibinfo {year}
  {2014})}\BibitemShut {NoStop}%
\bibitem [{Sup()}]{Supplemental}%
  \BibitemOpen
  \href@noop {} {}\bibinfo {note} {See Supplemental Material for experimental
  methods, additional characterization data and discussions}\BibitemShut
  {NoStop}%
\bibitem [{\citenamefont {Evans}\ \emph {et~al.}(2016)\citenamefont {Evans},
  \citenamefont {Sipahigil}, \citenamefont {Sukachev}, \citenamefont {Zibrov},\
  and\ \citenamefont {Lukin}}]{Evans2016}%
  \BibitemOpen
  \bibfield  {author} {\bibinfo {author} {\bibfnamefont {R.~E.}\ \bibnamefont
  {Evans}}, \bibinfo {author} {\bibfnamefont {A.}~\bibnamefont {Sipahigil}},
  \bibinfo {author} {\bibfnamefont {D.~D.}\ \bibnamefont {Sukachev}}, \bibinfo
  {author} {\bibfnamefont {A.~S.}\ \bibnamefont {Zibrov}},\ and\ \bibinfo
  {author} {\bibfnamefont {M.~D.}\ \bibnamefont {Lukin}},\ }\bibfield  {title}
  {\bibinfo {title} {Narrow-linewidth homogeneous optical emitters in diamond
  nanostructures via silicon ion implantation},\ }\href
  {https://doi.org/10.1103/PhysRevApplied.5.044010} {\bibfield  {journal}
  {\bibinfo  {journal} {Phys. Rev. Applied}\ }\textbf {\bibinfo {volume} {5}},\
  \bibinfo {pages} {044010} (\bibinfo {year} {2016})}\BibitemShut {NoStop}%
\bibitem [{\citenamefont {Hedrich}\ \emph {et~al.}(2020)\citenamefont
  {Hedrich}, \citenamefont {Rohner}, \citenamefont {Batzer}, \citenamefont
  {Maletinsky},\ and\ \citenamefont {Shields}}]{Hedrich2020}%
  \BibitemOpen
  \bibfield  {author} {\bibinfo {author} {\bibfnamefont {N.}~\bibnamefont
  {Hedrich}}, \bibinfo {author} {\bibfnamefont {D.}~\bibnamefont {Rohner}},
  \bibinfo {author} {\bibfnamefont {M.}~\bibnamefont {Batzer}}, \bibinfo
  {author} {\bibfnamefont {P.}~\bibnamefont {Maletinsky}},\ and\ \bibinfo
  {author} {\bibfnamefont {B.~J.}\ \bibnamefont {Shields}},\ }\bibfield
  {title} {\bibinfo {title} {Parabolic diamond scanning probes for single-spin
  magnetic field imaging},\ }\href
  {https://doi.org/10.1103/PhysRevApplied.14.064007} {\bibfield  {journal}
  {\bibinfo  {journal} {Phys. Rev. Applied}\ }\textbf {\bibinfo {volume}
  {14}},\ \bibinfo {pages} {064007} (\bibinfo {year} {2020})}\BibitemShut
  {NoStop}%
\bibitem [{\citenamefont {Thiering}\ and\ \citenamefont
  {Gali}(2018)}]{Thiering2018}%
  \BibitemOpen
  \bibfield  {author} {\bibinfo {author} {\bibfnamefont {G.}~\bibnamefont
  {Thiering}}\ and\ \bibinfo {author} {\bibfnamefont {A.}~\bibnamefont
  {Gali}},\ }\bibfield  {title} {\bibinfo {title} {Ab initio magneto-optical
  spectrum of group-iv vacancy color centers in diamond},\ }\href
  {https://doi.org/10.1103/PhysRevX.8.021063} {\bibfield  {journal} {\bibinfo
  {journal} {Phys. Rev. X}\ }\textbf {\bibinfo {volume} {8}},\ \bibinfo {pages}
  {021063} (\bibinfo {year} {2018})}\BibitemShut {NoStop}%
\bibitem [{\citenamefont {Farrer}(1969)}]{Farrer1969}%
  \BibitemOpen
  \bibfield  {author} {\bibinfo {author} {\bibfnamefont {R.~G.}\ \bibnamefont
  {Farrer}},\ }\bibfield  {title} {\bibinfo {title} {On the substitutional
  nitrogen donor in diamond},\ }\href
  {https://doi.org/10.1016/0038-1098(69)90593-6} {\bibfield  {journal}
  {\bibinfo  {journal} {Solid State Communications}\ }\textbf {\bibinfo
  {volume} {7}},\ \bibinfo {pages} {685} (\bibinfo {year} {1969})}\BibitemShut
  {NoStop}%
\bibitem [{\citenamefont {Chrenko}(1973)}]{Chrenko1973}%
  \BibitemOpen
  \bibfield  {author} {\bibinfo {author} {\bibfnamefont {R.~M.}\ \bibnamefont
  {Chrenko}},\ }\bibfield  {title} {\bibinfo {title} {{Boron, the dominant
  acceptor in semiconducting diamond}},\ }\href
  {https://doi.org/10.1103/PhysRevB.7.4560} {\bibfield  {journal} {\bibinfo
  {journal} {Physical Review B}\ }\textbf {\bibinfo {volume} {7}},\ \bibinfo
  {pages} {4560} (\bibinfo {year} {1973})}\BibitemShut {NoStop}%
\bibitem [{\citenamefont {Rogers}\ \emph {et~al.}(2019)\citenamefont {Rogers},
  \citenamefont {Wang}, \citenamefont {Liu}, \citenamefont {Antoniuk},
  \citenamefont {Osterkamp}, \citenamefont {Davydov}, \citenamefont {Agafonov},
  \citenamefont {Filipovski}, \citenamefont {Jelezko},\ and\ \citenamefont
  {Kubanek}}]{Rogers2019}%
  \BibitemOpen
  \bibfield  {author} {\bibinfo {author} {\bibfnamefont {L.~J.}\ \bibnamefont
  {Rogers}}, \bibinfo {author} {\bibfnamefont {O.}~\bibnamefont {Wang}},
  \bibinfo {author} {\bibfnamefont {Y.}~\bibnamefont {Liu}}, \bibinfo {author}
  {\bibfnamefont {L.}~\bibnamefont {Antoniuk}}, \bibinfo {author}
  {\bibfnamefont {C.}~\bibnamefont {Osterkamp}}, \bibinfo {author}
  {\bibfnamefont {V.~A.}\ \bibnamefont {Davydov}}, \bibinfo {author}
  {\bibfnamefont {V.~N.}\ \bibnamefont {Agafonov}}, \bibinfo {author}
  {\bibfnamefont {A.~B.}\ \bibnamefont {Filipovski}}, \bibinfo {author}
  {\bibfnamefont {F.}~\bibnamefont {Jelezko}},\ and\ \bibinfo {author}
  {\bibfnamefont {A.}~\bibnamefont {Kubanek}},\ }\bibfield  {title} {\bibinfo
  {title} {Single $\mathrm{Si}$-${V}^{\ensuremath{-}}$ centers in low-strain
  nanodiamonds with bulklike spectral properties and nanomanipulation
  capabilities},\ }\href {https://doi.org/10.1103/PhysRevApplied.11.024073}
  {\bibfield  {journal} {\bibinfo  {journal} {Phys. Rev. Applied}\ }\textbf
  {\bibinfo {volume} {11}},\ \bibinfo {pages} {024073} (\bibinfo {year}
  {2019})}\BibitemShut {NoStop}%
\bibitem [{\citenamefont {Shinotsuka}\ \emph {et~al.}(2015)\citenamefont
  {Shinotsuka}, \citenamefont {Tanuma}, \citenamefont {Powell},\ and\
  \citenamefont {Penn}}]{Shinotsuka2015}%
  \BibitemOpen
  \bibfield  {author} {\bibinfo {author} {\bibfnamefont {H.}~\bibnamefont
  {Shinotsuka}}, \bibinfo {author} {\bibfnamefont {S.}~\bibnamefont {Tanuma}},
  \bibinfo {author} {\bibfnamefont {C.~J.}\ \bibnamefont {Powell}},\ and\
  \bibinfo {author} {\bibfnamefont {D.~R.}\ \bibnamefont {Penn}},\ }\bibfield
  {title} {\bibinfo {title} {{Calculations of electron inelastic mean free
  paths. X. Data for 41 elemental solids over the 50 eV to 200 keV range with
  the relativistic full Penn algorithm}},\ }\href
  {https://doi.org/https://doi.org/10.1002/sia.5789} {\bibfield  {journal}
  {\bibinfo  {journal} {Surface and Interface Analysis}\ }\textbf {\bibinfo
  {volume} {47}},\ \bibinfo {pages} {871} (\bibinfo {year} {2015})}\BibitemShut
  {NoStop}%
\bibitem [{\citenamefont {Sangtawesin}\ \emph {et~al.}(2019)\citenamefont
  {Sangtawesin}, \citenamefont {Dwyer}, \citenamefont {Srinivasan},
  \citenamefont {Allred}, \citenamefont {Rodgers}, \citenamefont {De~Greve},
  \citenamefont {Stacey}, \citenamefont {Dontschuk}, \citenamefont {O'Donnell},
  \citenamefont {Hu}, \citenamefont {Evans}, \citenamefont {Jaye},
  \citenamefont {Fischer}, \citenamefont {Markham}, \citenamefont {Twitchen},
  \citenamefont {Park}, \citenamefont {Lukin},\ and\ \citenamefont
  {de~Leon}}]{Peace2019}%
  \BibitemOpen
  \bibfield  {author} {\bibinfo {author} {\bibfnamefont {S.}~\bibnamefont
  {Sangtawesin}}, \bibinfo {author} {\bibfnamefont {B.~L.}\ \bibnamefont
  {Dwyer}}, \bibinfo {author} {\bibfnamefont {S.}~\bibnamefont {Srinivasan}},
  \bibinfo {author} {\bibfnamefont {J.~J.}\ \bibnamefont {Allred}}, \bibinfo
  {author} {\bibfnamefont {L.~V.~H.}\ \bibnamefont {Rodgers}}, \bibinfo
  {author} {\bibfnamefont {K.}~\bibnamefont {De~Greve}}, \bibinfo {author}
  {\bibfnamefont {A.}~\bibnamefont {Stacey}}, \bibinfo {author} {\bibfnamefont
  {N.}~\bibnamefont {Dontschuk}}, \bibinfo {author} {\bibfnamefont {K.~M.}\
  \bibnamefont {O'Donnell}}, \bibinfo {author} {\bibfnamefont {D.}~\bibnamefont
  {Hu}}, \bibinfo {author} {\bibfnamefont {D.~A.}\ \bibnamefont {Evans}},
  \bibinfo {author} {\bibfnamefont {C.}~\bibnamefont {Jaye}}, \bibinfo {author}
  {\bibfnamefont {D.~A.}\ \bibnamefont {Fischer}}, \bibinfo {author}
  {\bibfnamefont {M.~L.}\ \bibnamefont {Markham}}, \bibinfo {author}
  {\bibfnamefont {D.~J.}\ \bibnamefont {Twitchen}}, \bibinfo {author}
  {\bibfnamefont {H.}~\bibnamefont {Park}}, \bibinfo {author} {\bibfnamefont
  {M.~D.}\ \bibnamefont {Lukin}},\ and\ \bibinfo {author} {\bibfnamefont
  {N.~P.}\ \bibnamefont {de~Leon}},\ }\bibfield  {title} {\bibinfo {title}
  {Origins of diamond surface noise probed by correlating single-spin
  measurements with surface spectroscopy},\ }\href
  {https://doi.org/10.1103/PhysRevX.9.031052} {\bibfield  {journal} {\bibinfo
  {journal} {Phys. Rev. X}\ }\textbf {\bibinfo {volume} {9}},\ \bibinfo {pages}
  {031052} (\bibinfo {year} {2019})}\BibitemShut {NoStop}%
\bibitem [{\citenamefont {Koslowski}\ \emph {et~al.}(1998)\citenamefont
  {Koslowski}, \citenamefont {Strobel}, \citenamefont {Wenig},\ and\
  \citenamefont {Ziemann}}]{Koslowski1998}%
  \BibitemOpen
  \bibfield  {author} {\bibinfo {author} {\bibfnamefont {B.}~\bibnamefont
  {Koslowski}}, \bibinfo {author} {\bibfnamefont {S.}~\bibnamefont {Strobel}},
  \bibinfo {author} {\bibfnamefont {M.~J.}\ \bibnamefont {Wenig}},\ and\
  \bibinfo {author} {\bibfnamefont {P.}~\bibnamefont {Ziemann}},\ }\bibfield
  {title} {\bibinfo {title} {{Roughness transitions of diamond(100) induced by
  hydrogen-plasma treatment}},\ }\href {https://doi.org/10.1007/s003390051318}
  {\bibfield  {journal} {\bibinfo  {journal} {Applied Physics A: Materials
  Science and Processing}\ }\textbf {\bibinfo {volume} {66}},\ \bibinfo {pages}
  {1159} (\bibinfo {year} {1998})}\BibitemShut {NoStop}%
\bibitem [{\citenamefont {Gaisinskaya}\ \emph {et~al.}(2009)\citenamefont
  {Gaisinskaya}, \citenamefont {Edrei}, \citenamefont {Hoffman},\ and\
  \citenamefont {Feldheim}}]{Gaisinskaya2009}%
  \BibitemOpen
  \bibfield  {author} {\bibinfo {author} {\bibfnamefont {A.}~\bibnamefont
  {Gaisinskaya}}, \bibinfo {author} {\bibfnamefont {R.}~\bibnamefont {Edrei}},
  \bibinfo {author} {\bibfnamefont {A.}~\bibnamefont {Hoffman}},\ and\ \bibinfo
  {author} {\bibfnamefont {Y.}~\bibnamefont {Feldheim}},\ }\bibfield  {title}
  {\bibinfo {title} {{Morphological evolution of polished single crystal (100)
  diamond surface exposed to microwave hydrogen plasma}},\ }\href
  {https://doi.org/10.1016/j.diamond.2009.09.014} {\bibfield  {journal}
  {\bibinfo  {journal} {Diamond and Related Materials}\ }\textbf {\bibinfo
  {volume} {18}},\ \bibinfo {pages} {1466} (\bibinfo {year}
  {2009})}\BibitemShut {NoStop}%
\bibitem [{\citenamefont {Crawford}\ \emph {et~al.}(2018)\citenamefont
  {Crawford}, \citenamefont {Tallaire}, \citenamefont {Li}, \citenamefont
  {Macdonald}, \citenamefont {Qi},\ and\ \citenamefont {Moran}}]{Crawford2018}%
  \BibitemOpen
  \bibfield  {author} {\bibinfo {author} {\bibfnamefont {K.~G.}\ \bibnamefont
  {Crawford}}, \bibinfo {author} {\bibfnamefont {A.}~\bibnamefont {Tallaire}},
  \bibinfo {author} {\bibfnamefont {X.}~\bibnamefont {Li}}, \bibinfo {author}
  {\bibfnamefont {D.~A.}\ \bibnamefont {Macdonald}}, \bibinfo {author}
  {\bibfnamefont {D.}~\bibnamefont {Qi}},\ and\ \bibinfo {author}
  {\bibfnamefont {D.~A.}\ \bibnamefont {Moran}},\ }\bibfield  {title} {\bibinfo
  {title} {{The role of hydrogen plasma power on surface roughness and carrier
  transport in transfer-doped H-diamond}},\ }\href
  {https://doi.org/10.1016/j.diamond.2018.03.005} {\bibfield  {journal}
  {\bibinfo  {journal} {Diamond and Related Materials}\ }\textbf {\bibinfo
  {volume} {84}},\ \bibinfo {pages} {48} (\bibinfo {year} {2018})}\BibitemShut
  {NoStop}%
\bibitem [{\citenamefont {Stacey}\ \emph {et~al.}(2012)\citenamefont {Stacey},
  \citenamefont {Karle}, \citenamefont {McGuinness}, \citenamefont {Gibson},
  \citenamefont {Ganesan}, \citenamefont {Tomljenovic‐Hanic}, \citenamefont
  {Greentree}, \citenamefont {Hoffman}, \citenamefont {Beausoleil},\ and\
  \citenamefont {Prawer}}]{Stacey2012}%
  \BibitemOpen
  \bibfield  {author} {\bibinfo {author} {\bibfnamefont {A.}~\bibnamefont
  {Stacey}}, \bibinfo {author} {\bibfnamefont {T.~J.}\ \bibnamefont {Karle}},
  \bibinfo {author} {\bibfnamefont {L.~P.}\ \bibnamefont {McGuinness}},
  \bibinfo {author} {\bibfnamefont {B.~C.}\ \bibnamefont {Gibson}}, \bibinfo
  {author} {\bibfnamefont {K.}~\bibnamefont {Ganesan}}, \bibinfo {author}
  {\bibfnamefont {S.}~\bibnamefont {Tomljenovic‐Hanic}}, \bibinfo {author}
  {\bibfnamefont {A.~D.}\ \bibnamefont {Greentree}}, \bibinfo {author}
  {\bibfnamefont {A.}~\bibnamefont {Hoffman}}, \bibinfo {author} {\bibfnamefont
  {R.~G.}\ \bibnamefont {Beausoleil}},\ and\ \bibinfo {author} {\bibfnamefont
  {S.}~\bibnamefont {Prawer}},\ }\bibfield  {title} {\bibinfo {title}
  {Depletion of nitrogen‐vacancy color centers in diamond via hydrogen
  passivation},\ }\href {https://doi.org/10.1063/1.3684612} {\bibfield
  {journal} {\bibinfo  {journal} {Applied Physics Letters}\ }\textbf {\bibinfo
  {volume} {100}},\ \bibinfo {pages} {071902} (\bibinfo {year}
  {2012})}\BibitemShut {NoStop}%
\bibitem [{\citenamefont {Edmonds}\ \emph {et~al.}(2008)\citenamefont
  {Edmonds}, \citenamefont {Newton}, \citenamefont {Martineau}, \citenamefont
  {Twitchen},\ and\ \citenamefont {Williams}}]{Edmonds2008}%
  \BibitemOpen
  \bibfield  {author} {\bibinfo {author} {\bibfnamefont {A.~M.}\ \bibnamefont
  {Edmonds}}, \bibinfo {author} {\bibfnamefont {M.~E.}\ \bibnamefont {Newton}},
  \bibinfo {author} {\bibfnamefont {P.~M.}\ \bibnamefont {Martineau}}, \bibinfo
  {author} {\bibfnamefont {D.~J.}\ \bibnamefont {Twitchen}},\ and\ \bibinfo
  {author} {\bibfnamefont {S.~D.}\ \bibnamefont {Williams}},\ }\bibfield
  {title} {\bibinfo {title} {Electron paramagnetic resonance studies of
  silicon-related defects in diamond},\ }\href
  {https://doi.org/10.1103/PhysRevB.77.245205} {\bibfield  {journal} {\bibinfo
  {journal} {Phys. Rev. B}\ }\textbf {\bibinfo {volume} {77}},\ \bibinfo
  {pages} {245205} (\bibinfo {year} {2008})}\BibitemShut {NoStop}%
\bibitem [{\citenamefont {Krivobok}\ \emph {et~al.}(2020)\citenamefont
  {Krivobok}, \citenamefont {Ekimov}, \citenamefont {Lyapin}, \citenamefont
  {Nikolaev}, \citenamefont {Skakov}, \citenamefont {Razgulov},\ and\
  \citenamefont {Kondrin}}]{Krivobok2020}%
  \BibitemOpen
  \bibfield  {author} {\bibinfo {author} {\bibfnamefont {V.~S.}\ \bibnamefont
  {Krivobok}}, \bibinfo {author} {\bibfnamefont {E.~A.}\ \bibnamefont
  {Ekimov}}, \bibinfo {author} {\bibfnamefont {S.~G.}\ \bibnamefont {Lyapin}},
  \bibinfo {author} {\bibfnamefont {S.~N.}\ \bibnamefont {Nikolaev}}, \bibinfo
  {author} {\bibfnamefont {Y.~A.}\ \bibnamefont {Skakov}}, \bibinfo {author}
  {\bibfnamefont {A.~A.}\ \bibnamefont {Razgulov}},\ and\ \bibinfo {author}
  {\bibfnamefont {M.~V.}\ \bibnamefont {Kondrin}},\ }\bibfield  {title}
  {\bibinfo {title} {{Observation of a 1.979-eV spectral line of a
  germanium-related color center in microdiamonds and nanodiamonds}},\ }\href
  {https://doi.org/10.1103/PhysRevB.101.144103} {\bibfield  {journal} {\bibinfo
   {journal} {Physical Review B}\ }\textbf {\bibinfo {volume} {101}},\ \bibinfo
  {pages} {144103} (\bibinfo {year} {2020})}\BibitemShut {NoStop}%
\bibitem [{\citenamefont {L{\"{u}}hmann}\ \emph {et~al.}(2020)\citenamefont
  {L{\"{u}}hmann}, \citenamefont {K{\"{u}}pper}, \citenamefont {Dietel},
  \citenamefont {Staacke}, \citenamefont {Meijer},\ and\ \citenamefont
  {Pezzagna}}]{Luhmann2020}%
  \BibitemOpen
  \bibfield  {author} {\bibinfo {author} {\bibfnamefont {T.}~\bibnamefont
  {L{\"{u}}hmann}}, \bibinfo {author} {\bibfnamefont {J.}~\bibnamefont
  {K{\"{u}}pper}}, \bibinfo {author} {\bibfnamefont {S.}~\bibnamefont
  {Dietel}}, \bibinfo {author} {\bibfnamefont {R.}~\bibnamefont {Staacke}},
  \bibinfo {author} {\bibfnamefont {J.}~\bibnamefont {Meijer}},\ and\ \bibinfo
  {author} {\bibfnamefont {S.}~\bibnamefont {Pezzagna}},\ }\bibfield  {title}
  {\bibinfo {title} {{Charge-State Tuning of Single SnV Centers in Diamond}},\
  }\href {https://doi.org/10.1021/acsphotonics.0c01123} {\bibfield  {journal}
  {\bibinfo  {journal} {ACS Photonics}\ }\textbf {\bibinfo {volume} {7}},\
  \bibinfo {pages} {3376} (\bibinfo {year} {2020})}\BibitemShut {NoStop}%
\bibitem [{\citenamefont {Sato}\ and\ \citenamefont {Kasu}(2013)}]{SATO201347}%
  \BibitemOpen
  \bibfield  {author} {\bibinfo {author} {\bibfnamefont {H.}~\bibnamefont
  {Sato}}\ and\ \bibinfo {author} {\bibfnamefont {M.}~\bibnamefont {Kasu}},\
  }\bibfield  {title} {\bibinfo {title} {{Maximum hole concentration for
  Hydrogen-terminated diamond surfaces with various surface orientations
  obtained by exposure to highly concentrated NO2}},\ }\href
  {https://doi.org/https://doi.org/10.1016/j.diamond.2012.10.007} {\bibfield
  {journal} {\bibinfo  {journal} {Diamond and Related Materials}\ }\textbf
  {\bibinfo {volume} {31}},\ \bibinfo {pages} {47} (\bibinfo {year}
  {2013})}\BibitemShut {NoStop}%
\bibitem [{\citenamefont {Kubovic}\ and\ \citenamefont
  {Kasu}(2010)}]{Kubovic2010}%
  \BibitemOpen
  \bibfield  {author} {\bibinfo {author} {\bibfnamefont {M.}~\bibnamefont
  {Kubovic}}\ and\ \bibinfo {author} {\bibfnamefont {M.}~\bibnamefont {Kasu}},\
  }\bibfield  {title} {\bibinfo {title} {Enhancement and stabilization of hole
  concentration of hydrogen-terminated diamond surface using ozone
  adsorbates},\ }\href {https://doi.org/10.1143/jjap.49.110208} {\bibfield
  {journal} {\bibinfo  {journal} {Japanese Journal of Applied Physics}\
  }\textbf {\bibinfo {volume} {49}},\ \bibinfo {pages} {110208} (\bibinfo
  {year} {2010})}\BibitemShut {NoStop}%
\bibitem [{\citenamefont {Strobel}\ \emph {et~al.}(2004)\citenamefont
  {Strobel}, \citenamefont {Riedel}, \citenamefont {Ristein},\ and\
  \citenamefont {Ley}}]{Strobel2004}%
  \BibitemOpen
  \bibfield  {author} {\bibinfo {author} {\bibfnamefont {P.}~\bibnamefont
  {Strobel}}, \bibinfo {author} {\bibfnamefont {M.}~\bibnamefont {Riedel}},
  \bibinfo {author} {\bibfnamefont {J.}~\bibnamefont {Ristein}},\ and\ \bibinfo
  {author} {\bibfnamefont {L.}~\bibnamefont {Ley}},\ }\bibfield  {title}
  {\bibinfo {title} {{Surface transfer doping of diamond}},\ }\href
  {https://doi.org/10.1038/nature02751} {\bibfield  {journal} {\bibinfo
  {journal} {Nature}\ }\textbf {\bibinfo {volume} {430}},\ \bibinfo {pages}
  {439} (\bibinfo {year} {2004})}\BibitemShut {NoStop}%
\bibitem [{\citenamefont {Russell}\ \emph {et~al.}(2013)\citenamefont
  {Russell}, \citenamefont {Cao}, \citenamefont {Qi}, \citenamefont {Tallaire},
  \citenamefont {Crawford}, \citenamefont {Wee},\ and\ \citenamefont
  {Moran}}]{Russell2013}%
  \BibitemOpen
  \bibfield  {author} {\bibinfo {author} {\bibfnamefont {S.~A.~O.}\
  \bibnamefont {Russell}}, \bibinfo {author} {\bibfnamefont {L.}~\bibnamefont
  {Cao}}, \bibinfo {author} {\bibfnamefont {D.}~\bibnamefont {Qi}}, \bibinfo
  {author} {\bibfnamefont {A.}~\bibnamefont {Tallaire}}, \bibinfo {author}
  {\bibfnamefont {K.~G.}\ \bibnamefont {Crawford}}, \bibinfo {author}
  {\bibfnamefont {A.~T.~S.}\ \bibnamefont {Wee}},\ and\ \bibinfo {author}
  {\bibfnamefont {D.~A.~J.}\ \bibnamefont {Moran}},\ }\bibfield  {title}
  {\bibinfo {title} {{Surface transfer doping of diamond by MoO3: A combined
  spectroscopic and Hall measurement study}},\ }\href
  {https://doi.org/10.1063/1.4832455} {\bibfield  {journal} {\bibinfo
  {journal} {Applied Physics Letters}\ }\textbf {\bibinfo {volume} {103}},\
  \bibinfo {pages} {202112} (\bibinfo {year} {2013})}\BibitemShut {NoStop}%
\end{thebibliography}%


\providecommand{\noopsort}[1]{}\providecommand{\singleletter}[1]{#1}%
\begin{thebibliography}{5}%
\makeatletter
\providecommand \@ifxundefined [1]{%
 \@ifx{#1\undefined}
}%
\providecommand \@ifnum [1]{%
 \ifnum #1\expandafter \@firstoftwo
 \else \expandafter \@secondoftwo
 \fi
}%
\providecommand \@ifx [1]{%
 \ifx #1\expandafter \@firstoftwo
 \else \expandafter \@secondoftwo
 \fi
}%
\providecommand \natexlab [1]{#1}%
\providecommand \enquote  [1]{``#1''}%
\providecommand \bibnamefont  [1]{#1}%
\providecommand \bibfnamefont [1]{#1}%
\providecommand \citenamefont [1]{#1}%
\providecommand \href@noop [0]{\@secondoftwo}%
\providecommand \href [0]{\begingroup \@sanitize@url \@href}%
\providecommand \@href[1]{\@@startlink{#1}\@@href}%
\providecommand \@@href[1]{\endgroup#1\@@endlink}%
\providecommand \@sanitize@url [0]{\catcode `\\12\catcode `\$12\catcode
  `\&12\catcode `\#12\catcode `\^12\catcode `\_12\catcode `\%12\relax}%
\providecommand \@@startlink[1]{}%
\providecommand \@@endlink[0]{}%
\providecommand \url  [0]{\begingroup\@sanitize@url \@url }%
\providecommand \@url [1]{\endgroup\@href {#1}{\urlprefix }}%
\providecommand \urlprefix  [0]{URL }%
\providecommand \Eprint [0]{\href }%
\providecommand \doibase [0]{https://doi.org/}%
\providecommand \selectlanguage [0]{\@gobble}%
\providecommand \bibinfo  [0]{\@secondoftwo}%
\providecommand \bibfield  [0]{\@secondoftwo}%
\providecommand \translation [1]{[#1]}%
\providecommand \BibitemOpen [0]{}%
\providecommand \bibitemStop [0]{}%
\providecommand \bibitemNoStop [0]{.\EOS\space}%
\providecommand \EOS [0]{\spacefactor3000\relax}%
\providecommand \BibitemShut  [1]{\csname bibitem#1\endcsname}%
\let\auto@bib@innerbib\@empty
\bibitem [{\citenamefont {Chu}\ \emph {et~al.}(2014)\citenamefont {Chu},
  \citenamefont {de~Leon}, \citenamefont {Shields}, \citenamefont {Hausmann},
  \citenamefont {Evans}, \citenamefont {Togan}, \citenamefont {Burek},
  \citenamefont {Markham}, \citenamefont {Stacey}, \citenamefont {Zibrov},
  \citenamefont {Yacoby}, \citenamefont {Twitchen}, \citenamefont {Loncar},
  \citenamefont {Park}, \citenamefont {Maletinsky},\ and\ \citenamefont
  {Lukin}}]{Chu2014}%
  \BibitemOpen
  \bibfield  {author} {\bibinfo {author} {\bibfnamefont {Y.}~\bibnamefont
  {Chu}}, \bibinfo {author} {\bibfnamefont {N.}~\bibnamefont {de~Leon}},
  \bibinfo {author} {\bibfnamefont {B.}~\bibnamefont {Shields}}, \bibinfo
  {author} {\bibfnamefont {B.}~\bibnamefont {Hausmann}}, \bibinfo {author}
  {\bibfnamefont {R.}~\bibnamefont {Evans}}, \bibinfo {author} {\bibfnamefont
  {E.}~\bibnamefont {Togan}}, \bibinfo {author} {\bibfnamefont {M.~J.}\
  \bibnamefont {Burek}}, \bibinfo {author} {\bibfnamefont {M.}~\bibnamefont
  {Markham}}, \bibinfo {author} {\bibfnamefont {A.}~\bibnamefont {Stacey}},
  \bibinfo {author} {\bibfnamefont {A.}~\bibnamefont {Zibrov}}, \bibinfo
  {author} {\bibfnamefont {A.}~\bibnamefont {Yacoby}}, \bibinfo {author}
  {\bibfnamefont {D.}~\bibnamefont {Twitchen}}, \bibinfo {author}
  {\bibfnamefont {M.}~\bibnamefont {Loncar}}, \bibinfo {author} {\bibfnamefont
  {H.}~\bibnamefont {Park}}, \bibinfo {author} {\bibfnamefont {P.}~\bibnamefont
  {Maletinsky}},\ and\ \bibinfo {author} {\bibfnamefont {M.}~\bibnamefont
  {Lukin}},\ }\href {https://doi.org/10.1021/nl404836p} {\bibfield  {journal}
  {\bibinfo  {journal} {Nano Letters}\ }\textbf {\bibinfo {volume} {14}},\
  \bibinfo {pages} {1982} (\bibinfo {year} {2014})}\BibitemShut {NoStop}%
\bibitem [{\citenamefont {Evans}\ \emph {et~al.}(2016)\citenamefont {Evans},
  \citenamefont {Sipahigil}, \citenamefont {Sukachev}, \citenamefont {Zibrov},\
  and\ \citenamefont {Lukin}}]{Evans2016}%
  \BibitemOpen
  \bibfield  {author} {\bibinfo {author} {\bibfnamefont {R.~E.}\ \bibnamefont
  {Evans}}, \bibinfo {author} {\bibfnamefont {A.}~\bibnamefont {Sipahigil}},
  \bibinfo {author} {\bibfnamefont {D.~D.}\ \bibnamefont {Sukachev}}, \bibinfo
  {author} {\bibfnamefont {A.~S.}\ \bibnamefont {Zibrov}},\ and\ \bibinfo
  {author} {\bibfnamefont {M.~D.}\ \bibnamefont {Lukin}},\ }\href
  {https://doi.org/10.1103/PhysRevApplied.5.044010} {\bibfield  {journal}
  {\bibinfo  {journal} {Phys. Rev. Applied}\ }\textbf {\bibinfo {volume} {5}},\
  \bibinfo {pages} {044010} (\bibinfo {year} {2016})}\BibitemShut {NoStop}%
\bibitem [{\citenamefont {Hedrich}\ \emph {et~al.}(2020)\citenamefont
  {Hedrich}, \citenamefont {Rohner}, \citenamefont {Batzer}, \citenamefont
  {Maletinsky},\ and\ \citenamefont {Shields}}]{Hedrich2020}%
  \BibitemOpen
  \bibfield  {author} {\bibinfo {author} {\bibfnamefont {N.}~\bibnamefont
  {Hedrich}}, \bibinfo {author} {\bibfnamefont {D.}~\bibnamefont {Rohner}},
  \bibinfo {author} {\bibfnamefont {M.}~\bibnamefont {Batzer}}, \bibinfo
  {author} {\bibfnamefont {P.}~\bibnamefont {Maletinsky}},\ and\ \bibinfo
  {author} {\bibfnamefont {B.~J.}\ \bibnamefont {Shields}},\ }\href
  {https://doi.org/10.1103/PhysRevApplied.14.064007} {\bibfield  {journal}
  {\bibinfo  {journal} {Phys. Rev. Applied}\ }\textbf {\bibinfo {volume}
  {14}},\ \bibinfo {pages} {064007} (\bibinfo {year} {2020})}\BibitemShut
  {NoStop}%
\bibitem [{\citenamefont {Sangtawesin}\ \emph {et~al.}(2019)\citenamefont
  {Sangtawesin}, \citenamefont {Dwyer}, \citenamefont {Srinivasan},
  \citenamefont {Allred}, \citenamefont {Rodgers}, \citenamefont {De~Greve},
  \citenamefont {Stacey}, \citenamefont {Dontschuk}, \citenamefont {O'Donnell},
  \citenamefont {Hu}, \citenamefont {Evans}, \citenamefont {Jaye},
  \citenamefont {Fischer}, \citenamefont {Markham}, \citenamefont {Twitchen},
  \citenamefont {Park}, \citenamefont {Lukin},\ and\ \citenamefont
  {de~Leon}}]{Peace2019}%
  \BibitemOpen
  \bibfield  {author} {\bibinfo {author} {\bibfnamefont {S.}~\bibnamefont
  {Sangtawesin}}, \bibinfo {author} {\bibfnamefont {B.~L.}\ \bibnamefont
  {Dwyer}}, \bibinfo {author} {\bibfnamefont {S.}~\bibnamefont {Srinivasan}},
  \bibinfo {author} {\bibfnamefont {J.~J.}\ \bibnamefont {Allred}}, \bibinfo
  {author} {\bibfnamefont {L.~V.~H.}\ \bibnamefont {Rodgers}}, \bibinfo
  {author} {\bibfnamefont {K.}~\bibnamefont {De~Greve}}, \bibinfo {author}
  {\bibfnamefont {A.}~\bibnamefont {Stacey}}, \bibinfo {author} {\bibfnamefont
  {N.}~\bibnamefont {Dontschuk}}, \bibinfo {author} {\bibfnamefont {K.~M.}\
  \bibnamefont {O'Donnell}}, \bibinfo {author} {\bibfnamefont {D.}~\bibnamefont
  {Hu}}, \bibinfo {author} {\bibfnamefont {D.~A.}\ \bibnamefont {Evans}},
  \bibinfo {author} {\bibfnamefont {C.}~\bibnamefont {Jaye}}, \bibinfo {author}
  {\bibfnamefont {D.~A.}\ \bibnamefont {Fischer}}, \bibinfo {author}
  {\bibfnamefont {M.~L.}\ \bibnamefont {Markham}}, \bibinfo {author}
  {\bibfnamefont {D.~J.}\ \bibnamefont {Twitchen}}, \bibinfo {author}
  {\bibfnamefont {H.}~\bibnamefont {Park}}, \bibinfo {author} {\bibfnamefont
  {M.~D.}\ \bibnamefont {Lukin}},\ and\ \bibinfo {author} {\bibfnamefont
  {N.~P.}\ \bibnamefont {de~Leon}},\ }\href
  {https://doi.org/10.1103/PhysRevX.9.031052} {\bibfield  {journal} {\bibinfo
  {journal} {Phys. Rev. X}\ }\textbf {\bibinfo {volume} {9}},\ \bibinfo {pages}
  {031052} (\bibinfo {year} {2019})}\BibitemShut {NoStop}%
\bibitem [{\citenamefont {D'Haenens-Johansson}\ \emph
  {et~al.}(2011)\citenamefont {D'Haenens-Johansson}, \citenamefont {Edmonds},
  \citenamefont {Green}, \citenamefont {Newton}, \citenamefont {Davies},
  \citenamefont {Martineau}, \citenamefont {Khan},\ and\ \citenamefont
  {Twitchen}}]{Ulrika2011}%
  \BibitemOpen
  \bibfield  {author} {\bibinfo {author} {\bibfnamefont {U.~F.~S.}\
  \bibnamefont {D'Haenens-Johansson}}, \bibinfo {author} {\bibfnamefont
  {A.~M.}\ \bibnamefont {Edmonds}}, \bibinfo {author} {\bibfnamefont {B.~L.}\
  \bibnamefont {Green}}, \bibinfo {author} {\bibfnamefont {M.~E.}\ \bibnamefont
  {Newton}}, \bibinfo {author} {\bibfnamefont {G.}~\bibnamefont {Davies}},
  \bibinfo {author} {\bibfnamefont {P.~M.}\ \bibnamefont {Martineau}}, \bibinfo
  {author} {\bibfnamefont {R.~U.~A.}\ \bibnamefont {Khan}},\ and\ \bibinfo
  {author} {\bibfnamefont {D.~J.}\ \bibnamefont {Twitchen}},\ }\href
  {https://doi.org/10.1103/PhysRevB.84.245208} {\bibfield  {journal} {\bibinfo
  {journal} {Phys. Rev. B}\ }\textbf {\bibinfo {volume} {84}},\ \bibinfo
  {pages} {245208} (\bibinfo {year} {2011})}\BibitemShut {NoStop}%
\end{thebibliography}%

\end{document}


\title{Supplemental Material for \\``Neutral silicon vacancy centers in undoped diamond via surface control''}
\author{Zi-Huai Zhang}
\affiliation{%
Department of Electrical and Computer Engineering, Princeton University, Princeton, New Jersey 08544, USA
}%

\author{Josh A. Zuber}
\affiliation{
Department of Physics, University of Basel, Klingelbergstrasse 82, 4056 Basel, Switzerland
}
\affiliation{Swiss Nanoscience Institute, Klingelbergstrasse 82, 4056 Basel, Switzerland}

\author{Lila V. H. Rodgers}
\affiliation{%
Department of Electrical and Computer Engineering, Princeton University, Princeton, New Jersey 08544, USA
}%

\author{Xin Gui}
\affiliation{
Department of Chemistry, Princeton University, Princeton, New Jersey 08544, USA
}

\author{Paul Stevenson}
\affiliation{
Department of Physics, Northeastern University, Boston, Massachusetts 02115, USA
}

\author{Minghao Li}
\affiliation{
Department of Physics, University of Basel, Klingelbergstrasse 82, 4056 Basel, Switzerland
}
\affiliation{Swiss Nanoscience Institute, Klingelbergstrasse 82, 4056 Basel, Switzerland}

\author{Marietta Batzer}
\affiliation{
Department of Physics, University of Basel, Klingelbergstrasse 82, 4056 Basel, Switzerland
}
\affiliation{Swiss Nanoscience Institute, Klingelbergstrasse 82, 4056 Basel, Switzerland}

\author{Marcel.li Grimau}
\affiliation{
Department of Physics, University of Basel, Klingelbergstrasse 82, 4056 Basel, Switzerland
}
\affiliation{Swiss Nanoscience Institute, Klingelbergstrasse 82, 4056 Basel, Switzerland}

\author{Brendan Shields}
\affiliation{
Department of Physics, University of Basel, Klingelbergstrasse 82, 4056 Basel, Switzerland
}
\affiliation{Swiss Nanoscience Institute, Klingelbergstrasse 82, 4056 Basel, Switzerland}

\author{Andrew M.
Edmonds}
\affiliation{
Element Six, Harwell, OX11 0QR, United Kingdom
}

\author{Nicola Palmer}
\affiliation{
Element Six, Harwell, OX11 0QR, United Kingdom
}

\author{Matthew L. Markham}
\affiliation{
Element Six, Harwell, OX11 0QR, United Kingdom
}

\author{Robert J. Cava}
\affiliation{
Department of Chemistry, Princeton University, Princeton, New Jersey 08544, USA
}

\author{Patrick Maletinsky}
\affiliation{
Department of Physics, University of Basel, Klingelbergstrasse 82, 4056 Basel, Switzerland
}
\affiliation{Swiss Nanoscience Institute, Klingelbergstrasse 82, 4056 Basel, Switzerland}

\author{Nathalie P. de Leon}%
 \email{npdeleon@princeton.edu}
\affiliation{%
Department of Electrical and Computer Engineering, Princeton University, Princeton, New Jersey 08544, USA
}%

\date{\today}
\maketitle
\label{Sec:SI}

\setcounter{figure}{0}
\setcounter{section}{0}
\renewcommand{\thefigure}{S\arabic{figure}}
\renewcommand{\thetable}{\Roman{table}}
\renewcommand{\thesection}{\Roman{section}}

\section{\label{SI_Setup} SUPPLEMENTARY EXPERIMENTAL METHODS}
\subsection{Experimental setup}
Photoluminescence (PL) measurements of neutral silicon vacancy (\siv{}) centers and negatively charged silicon vacancy (\sivm{}) centers were performed in two different home-built confocal microscopes. 

For \siv{}, measurements were conducted in a cryogenic confocal microscope optimized for near infrared (NIR) with a Janis helium flow cryostat and a 0.65 NA, 50X objective (Olympus LCPLN50XIR). The excitation channel and detection channel were combined with a 925~nm dichroic beam splitter (Semrock FF925-Di01-25-D). The resonant excitation channel was combined with the detection channel with a 10/90 beamspliter (Thorlabs BS044). The set temperature of the cryostat was $\sim$10~K unless otherwise noted. Emission spectrum measurements and photoluminescence excitation (PLE) spectroscopy of the bound exciton transition were performed using off-resonant excitation with a tunable diode laser (Toptica DL pro 850 nm). The detection channel was filtered using a 937 nm long pass filter (Semrock FF01-937/LP-25). The fluorescence signal was collected into a fiber and sent to a CCD spectrometer (Princeton Instruments Acton SP-2300i with Pixis 100 CCD and 300 g/mm grating). PLE of the zero-phonon line (ZPL) and optically detected magnetic resonance (ODMR) were performed using resonant excitation with a tunable diode laser (Toptica CTL 950) with the sideband emission of \siv{} filtered by a 980 nm long pass filder (Semrock LP02-980RE-25) and detected by a superconducting nanowire detector (Quantum Opus, optimized for 950 - 1100 nm). For ODMR, microwave (MW) excitation was applied using a thin wire stretched across the sample. The MW excitation was generated with a signal generator (Keysight N9310A) and then amplified by a high-power MW amplifier (Triad TB1003).
Two 0.8 - 2 GHz MW circulators (Ditom D3C0802S) were added after the amplifier for circuit protection. The MW
excitation was pulsed using a fast MW switch (Mini-Circuits ZASWA-2-50DR+) gated by a TTL pulse generator (Spincore PBESR-PRO-500-PCI). For ODMR, the MW excitation was modulated to have a 2~ms period with 50\% duty cycle.

For \sivm{}, measurements were conducted in a confocal microscope optimized for visible wavelengths under ambient conditions. PL measurements were performed with a 561 nm laser (Cobolt 08-01-DPL). A 0.9 NA, 100X objective (Olympus MPlanFL N 100X) was used to focus the light onto the sample. The excitation channel and
detection channel were combined with a 635 nm dichroic beam splitter (Semrock Di03-R635-t3-25-D). The detection channel was filtered
using a 647 nm long pass filter (Semrock BLP01-647R-25). Emission spectrum was again measured using the CCD spectrometer. 

The surface analysis measurements were performed at the Imaging and Analysis Center at Princeton University. X-ray photoelectron spectroscopy (XPS) was performed
with a Thermo Fisher K-Alpha spectrometer, collecting photoelectrons normal to the surface. Atomic force microscopy (AFM) was performed with either a Bruker ICON3 AFM or a Bruker NanoMan AFM operating in ac tapping mode (AFM tip Asylum Research AC160TS-R3, resonance frequency 300 kHz).

\subsection{Diamond fabrication methods}
\begin{figure}[h!]
  \centering
  \includegraphics[width = 129mm]{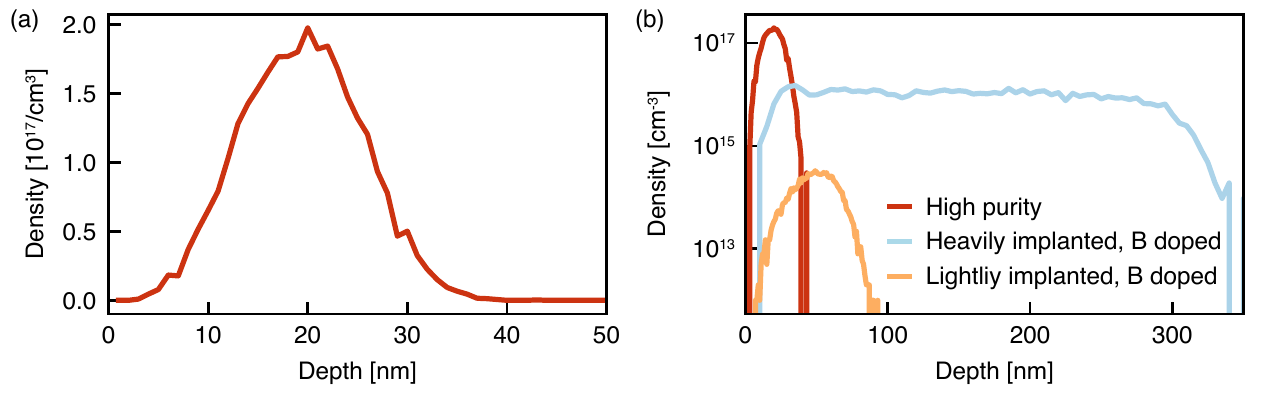}
  \caption{(a) Depth profile of the implanted Si ions calculated using SRIM in the high purity diamond. (b) Depth profile of the implanted Si ions calculated using SRIM for the high purity diamond as well as the two boron doped diamonds.}
  \label{fig:FigSRIM}
\end{figure}

A high purity diamond grown by plasma chemical vapor deposition (Element Six ``electronic grade") was used in the experiments. The diamond is nominally undoped with nitrogen concentration below 5 ppb and boron concentration below 1 ppb. The diamond was scaife polished \
into a 50 $\mu$m membrane, and subsequently etched with Ar/Cl$_2$ and O$_2$ to remove subsurface strain \cite{Chu2014}. The diamond was implanted with $^{28}$Si at 25~keV with total fluence of 3$\times$ $10^{11}$ cm$^{-2}$ at 7$\degree$ tilt angle. We simulate the distribution of implanted Si ions using stopping range in matter (SRIM) (Fig.~\ref{fig:FigSRIM}(a)), and estimate the average depth to be about 20~nm.

Following ion implantation, we cleaned the sample in a tri-acid mixture and performed high-temperature high-vacuum annealing to the sample in a home-built vacuum annealing oven with a heater controller (Tectra HC3500) and a heater plate (BORALECTRIC). The annealing steps are 400$\degree$C for 4 hours, 800$\degree$C for 11 hours and 1100$\degree$C for 2 hours \cite{Evans2016}. After annealing, parabolic reflectors (PR) of approximately 300~nm diameter were fabricated using a plasma etcher (Sentech ICP-RIE SI 500) to enhance the collection efficiency following the etch recipe in Ref.~\cite{Hedrich2020}.

The bulk doped \siv{} containing sample shown in Fig.~4(a) and Fig.~4(b) was grown by chemical vapor deposition (Element Six) and doped with silicon during growth. The resulting \siv{} concentration in the sample was estimated to be 5.6$\times10^{15}$ cm$^{-3}$ based on absorption measurement.

Two additional diamonds are discussed in the supplemental material as reference samples for comparing surface transfer doping and bulk doping. The two diamonds are boron doped diamonds from DiamFab, specified with [B] $\sim 10^{17}$ cm$^{-3}$ and [N] below $1.8\times10^{14}$ cm$^{-3}$. The first boron doped diamond (heavily implanted) was implanted with silicon using multiple energies with a total fluence of 3$\times10^{11}$ cm$^{-2}$. The implantation recipe is:

\begin{center}
    \begin{tabular}{c|c|c|c}
     Step\# & Energy (keV)    &  Dose (cm$^{-2}$) & Tilt (degrees)\\
     \hline
      1 & 400  &  6.83$\times10^{10}$ & 7\\
      2 &310  &  5.24$\times10^{10}$ & 7\\
      3 &240  &  4.67$\times10^{10}$ & 7\\
      4 &180   &  3.98$\times10^{10}$ & 7\\
      5 &120  &  3.59$\times10^{10}$ & 7\\
      6 &80  &  2.85$\times10^{10}$& 7 \\
      7 &40  &  2.85$\times10^{10}$ & 7
      
    \end{tabular}
\end{center} 
The second boron doped diamond (lightly implanted) was silicon implanted with a fluence of 1$\times10^{9}$ cm$^{-2}$ at 70~keV. After implantation, high temperature high vacuum annealing was performed in a Lindberg Blue tube furnace with the following steps \cite{Peace2019}: (1) Ramp to 100 \degree C over 1 hour, hold for 11 hours;
(2) Ramp to 400 \degree C over 4 hours, hold for 8 hours; (3) Ramp to 800 \degree C over 6 hours, hold for 8 hours; (4) Ramp to 1200 \degree C over 6 hours, hold for 2 hours; (4) Let cool to room temperature. The pressure over the whole annealing was less than $2 \times 10^{-6}$ Torr. After the initial annealing, the heavily implanted sample was annealed for a second time at 1850 \degree C for 10~min under stabilizing pressure to further reduce implantation damage.

The distributions of implanted Si ions for the two boron doped diamonds are simulated using SRIM, as shown in Fig.~\ref{fig:FigSRIM}(b).

\subsection{Surface preparation and characterization}
SiV centers were characterized under different surfaces throughout the main text. For reversible tuning of SiV, we toggled the surface between H-termination and O-termination. For ODMR measurements, the surface was cleaned in piranha solution (a 1:2 mixture of hydrogen peroxide in concentrated sulfuric acid) 3 times after H termination. 

For O-terminated surfaces, we prepare the surface by cleaning the sample in a refluxing 1:1:1 mixture of concentrated sulfuric, nitric, and perchloric acids (tri-acid
cleaning) for 2 hours. 

In order to prepare the H-terminated surface, we first clean the sample with tri-acid to ensure the sample is contamination free and then anneal the sample in either pure hydrogen or forming gas (5\% H$_2$ and 95\% Ar). For hydrogen annealing, the sample was annealed for 6 hours with a hydrogen flow rate of 40 sccm in a tube furnace at 750$\degree$C and subsequently cooled to room temperature under forming gas with flow rate of 200 sccm. For forming gas annealing, the annealing was performed in a tube furnace (Lindberg
Blue Mini-Mite with high-purity quartz process tube) under
continuous flow of forming gas at atmospheric pressure. The annealing steps are: (1) Ramp to 100 \degree C over 1 hour, hold for 2 hours; (2) Ramp to 800 \degree C over 2.5 hours, hold for 72 hours; (3) Let cool to room temperature inside the furnace. After surface preparation, the diamond membrane was attached to a silicon chip using Crystalbond 509 with the unetched surface facing up, and the whole silicon chip was mounted on a gold plated copper sample mount using GE varnish (CMR direct) for measurements. We note that initial observation of \siv{} was under an H-terminated surface prepared by hydrogen annealing followed by exposure to ambient conditions for over 1.5 years. The  data shown in the main text and supplemental materials are with H-termination prepared using forming gas annealing unless otherwise noted. 

\begin{figure}[h!]
  \centering
  \includegraphics[width = 86mm]{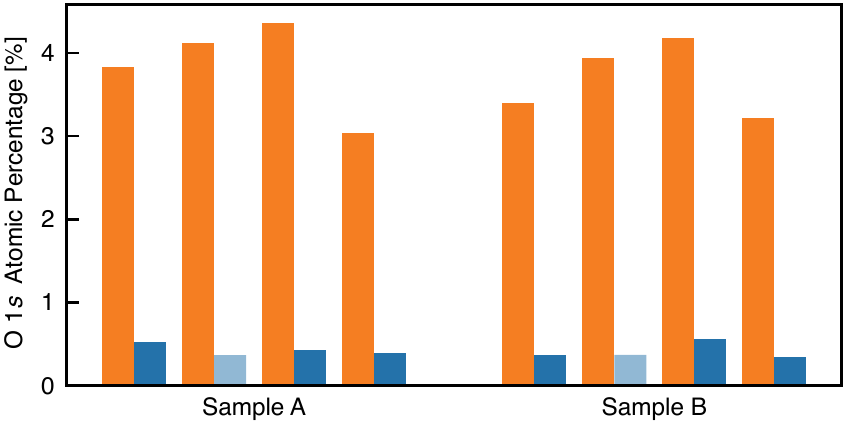}
  \caption{Repetitive forming gas annealing and tri-acid cleaning on two test samples. The surface termination was characterized by the oxygen~1$s$ atomic percentage where the orange bars denote surfaces after triacid cleaning and the blue bars denote surfaces after forming gas annealing. The hold time at 800 \degree C was 24 hours for the light blue bar while the hold time was 72 hours for the rest of tests. }
  \label{fig:FigS1}
\end{figure}

The surface terminations after different processing were characterized by X-ray photoelectron spectroscopy (XPS). We observed consistent reduction of the oxygen 1$s$ peak in XPS after forming gas annealing. After tri-acid cleaning, the oxygen 1$s$ peak also recovers to a value similar to that before H-termination. In Fig.~\ref{fig:FigS1}, we show XPS characterization of two test samples after repetitive triacid cleaning and forming gas annealing. The oxygen atomic percentage toggles consistently, demonstrating the reversible and non-destructive nature of the processes. 

AFM scans were measured on test diamonds before and after the forming gas annealing. In Fig.~2(c), two \mbox{500~nm $\times$ 500~nm} scan were shown to compare the surfaces. The scan image before annealing was filtered using the python package pystripe to remove periodic instrumental stripe noises. 

\section{\label{MoreData} Additional Measurements of Silicon vacancy centers}

\subsection{Initial observation of \siv{} under H-termination after long-term air exposure}
\siv{} centers were first observed in the sample under a surface prepared by H-termination via hydrogen annealing and subsequent long-term ($\sim$1.5 years) air exposure. The \siv{} centers prepared under H-termination via hydrogen annealing showed qualitatively similar behavior as the \siv{} centers prepared under H-termination via forming gas annealing, even though the former surface was subject to long-term air exposure. Therefore, we conclude that the surface condition was long-lived and robust to air exposure. The accumulation of surface adsorbates complicates quantification of the H-termination after long-term air exposure because adventitious carbon, water, and other adsorbates contain oxygen.

\begin{figure}[h!]
  \centering
  \includegraphics[width = 86mm]{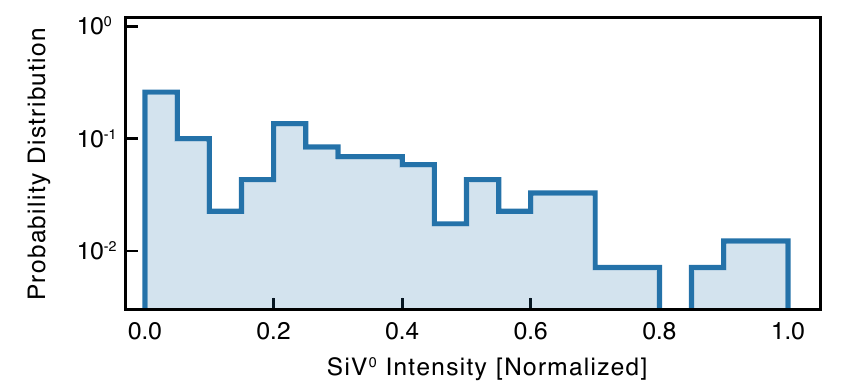}
  \caption{\siv{} emission intensity distribution under H-termination prepared by hydrogen annealing with subsequent long term air exposure. The measurement was performed with a set temperature of 10~K using $\sim$11 mW 857~nm excitation. Qualitative comparison between \siv{} under this surface condition and \siv{} prepared under fresh H-termination via forming gas annealing is not provided here due to changes in sample mounting scheme.}
  \label{fig:Tdep}
\end{figure}

\subsection{Temperature dependence}
The ZPL emission of \siv{} was previously shown to be highly temperature dependent, where population trapping into a dark state resulted in quenching of ZPL emission below 70~K \cite{Ulrika2011}. In our sample under H-termination, we performed temperature dependent PL measurements and observed a similar quenching effect, as shown in Fig.~\ref{fig:Tdep}. The trend of temperature dependence agrees with previous reports where a maximum of PL emission into 946 nm ZPL is reached at around 70K.

\begin{figure}[h!]
  \centering
  \includegraphics[width = 86mm]{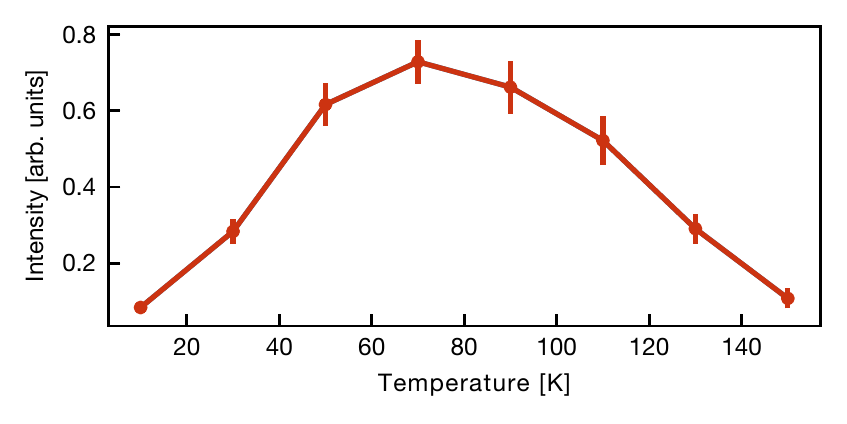}
  \caption{Temperature dependence of 946 nm ZPL emission intensity of \siv{} under H-termination. The data is an average over measurements of 36 PRs.}
  \label{fig:Tdep}
\end{figure}

\begin{figure}[h!]
  \centering
  \includegraphics[width = 86mm]{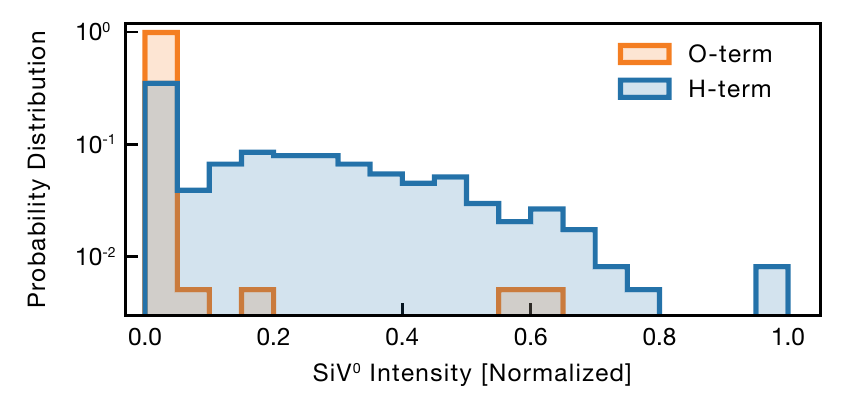}
  \caption{\siv{} emission intensity distribution under H- and O-termination measured at 70 K with $\sim$11 mW 857 nm excitation.}
  \label{fig:70K_Histogram}
\end{figure}

We performed complementary measurements of the emission statistics of \siv{} at 70~K under different surfaces, as shown in Fig.~\ref{fig:70K_Histogram}. The result agrees with the measurements at 10~K shown in the main text. 

\subsection{Observation of \siv{} in unetched regions and comparison to \siv{} formed in boron doped diamonds}
\begin{figure}[h!]
  \centering
  \includegraphics[width = 129mm]{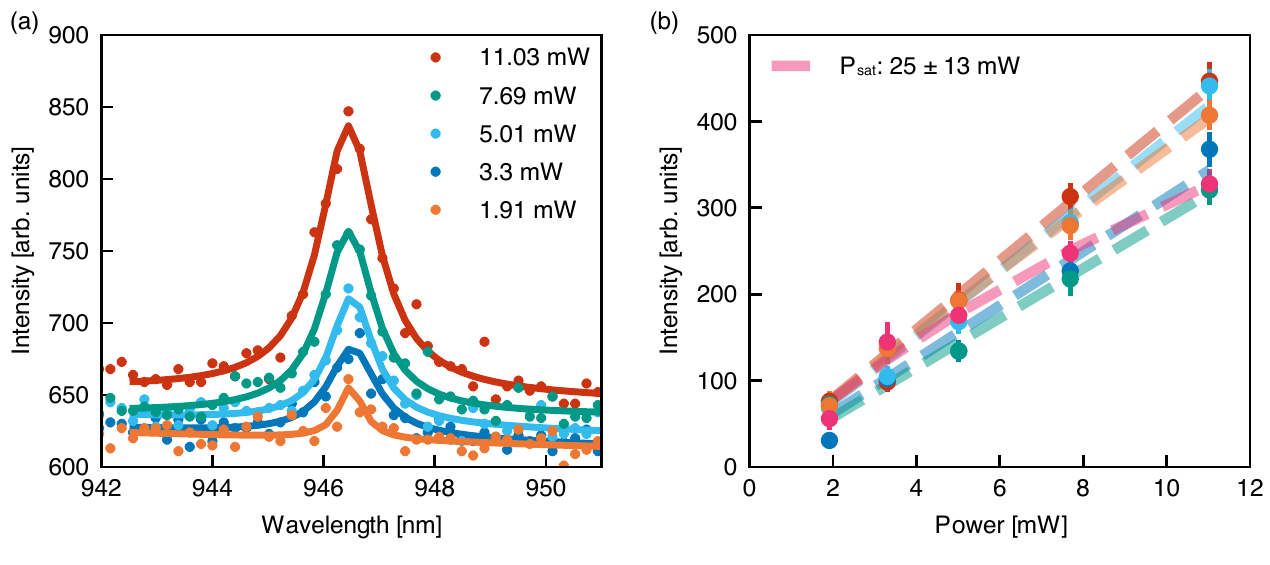}
  \caption{\siv{} emission in unetched region under H-termination. (a) Emission spectra from a random position with different excitation powers. Solid curves show Lorentzian fits to the data. The amplitudes extracted from the fits are used in (b). (b) Saturation study at 6 random positions. Dashed lines show fits of the emission intensity using $I(P) = I_s/(1 + P_0/P)$ where $P$ is the optical power, $P_0$ is the saturation power and $I_s$ is the saturated emission intensity. Only one of the positions showed some signature of saturation. The fit results for other positions were consistent with a linear dependence. Measurements were performed at 70~K with 857 nm excitation.}
  \label{fig:FigS3}
\end{figure}

In the main text, characterization of \siv{} emission was done in the etched PRs. Here, we show \siv{} measurements on the unetched region under H-termination. Emission spectra from a randomly chosen position is shown in Fig.~\ref{fig:FigS3}(a). We performed power dependent measurements at 6 random positions in a 40~$\mu$m $\times$ 40~$\mu$m region, and intensity from only one of the positions showed some sign of saturation, as shown in Fig.~\ref{fig:FigS3}(b). The ability to observe similar levels of \siv{} emission at different positions suggests that the \siv{} density under H-termination was high enough to form a uniform layer.

\begin{figure}[h!]
  \centering
  \includegraphics[width = 129mm]{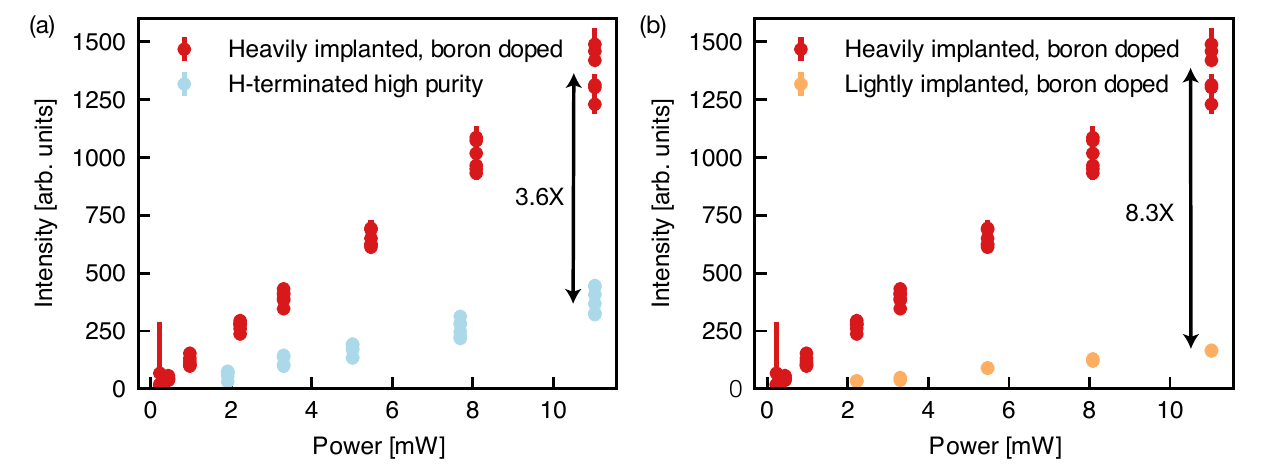}
  \caption{Comparison of \siv{} emission in the H-terminated high purity diamond and boron doped diamonds. (a) Emission intensity comparison between the H-terminated high purity diamond and the heavily implanted boron doped diamond. The intensities were measured at a few different positions on the samples at different powers. The heavily implanted boron doped diamond showed a factor of 3.6 higher \siv{} emission compared to the high purity diamond under H-termination.  (b) Emission intensity comparison between the heavily implanted and the lightly implanted boron doped diamonds. The intensities were measured on a few different positions on the samples at different powers. The heavily implanted boron doped diamond only showed a factor of 8.3 higher \siv{} emission compared to the lightly implanted boron doped diamond though the implantation fluence was higher by a factor of 300. Measurements were performed at 70~K with 857~nm excitation.}
  \label{fig:FigSample}
\end{figure}

We compare the emission intensity of \siv{} formed under an H-terminated surface to the emission intensity of \siv{} prepared in boron doped diamonds. We observe that the \siv{} intensity was within a factor of four of the \siv{} intensity from the heavily implanted boron doped diamond (Fig.~\ref{fig:FigSample}(a)). Discrepancy of normalized \siv{} emission intensity among boron doped diamonds precludes a straightforward quantitative estimate of the total conversion efficiency. For example, even though the total implantation fluence was 300 times higher for the heavily implanted boron doped diamond, the intensity of \siv{} was only a factor of 8.3 higher compared to the lightly implanted sample (Fig.~\ref{fig:FigSample}(b)). Such a discrepancy could arise from density dependent quenching of \siv{} emission. Nevertheless, the fact that the H-terminated sample (with highest implantation density) only showed a factor of 3.6 lower signal than the heavily implanted boron doped sample demonstrated that surface transfer doping was comparably effective to bulk doping. 

\subsection{Surface stability after iterative piranha cleanings}
\begin{figure}[h!]
  \centering
  \includegraphics[width = 129mm]{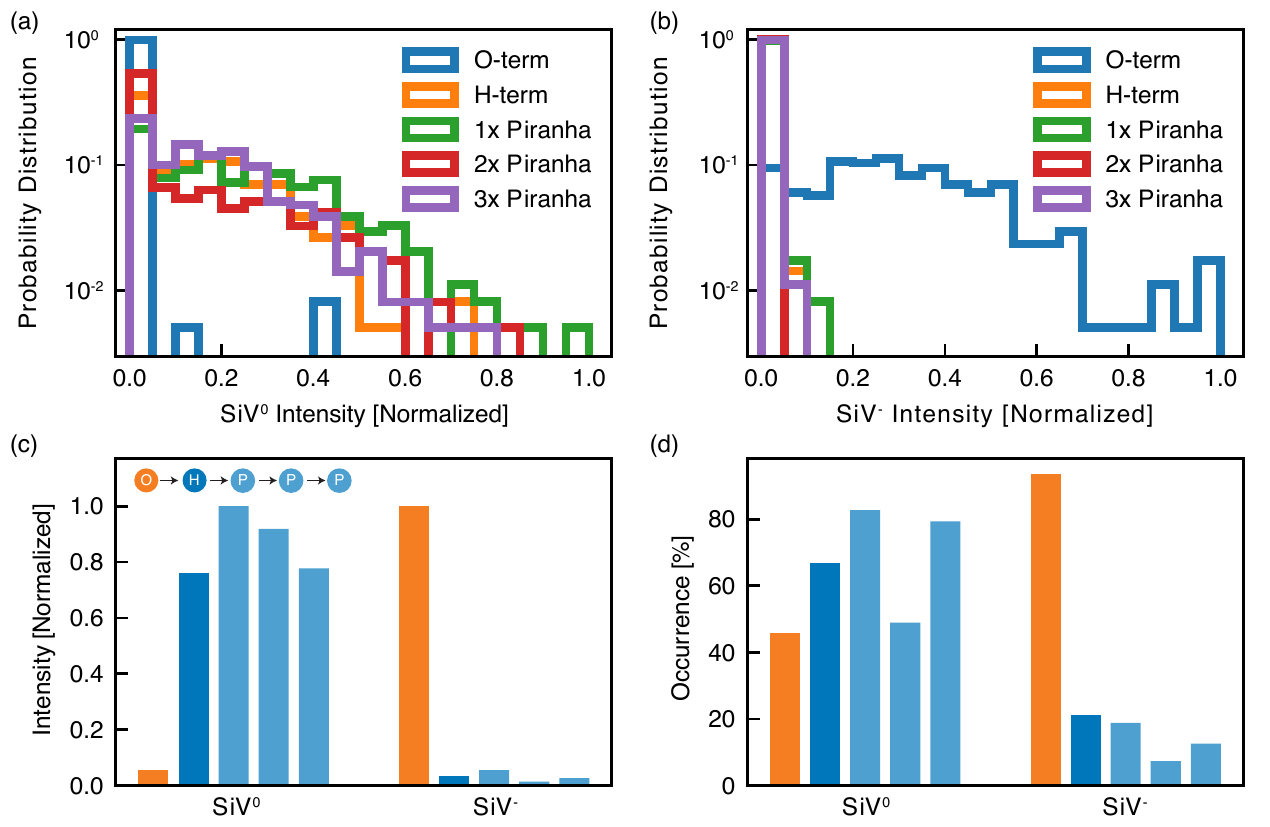}
  \caption{Effect of piranha cleaning on SiV emission under an H-terminated surface. (a) Histograms of \siv{} emission under different surfaces. Three piranha cleaning steps were performed after H-termination. Histogram under O-termination is shown for comparison. (b) Histograms of \sivm{} emission under the same surface conditions. Histogram under O-termination is shown for comparison. (c) Average emission intensity of \siv{} and \sivm{} under different surfaces. Inset: order of surface conditions where O denotes O-termination, H denotes H-termination, and P denotes H-terminated surface after piranha cleaning. (d) Occurrence of \siv{} and \sivm{} under difference surfaces. Measurements for \siv{} were performed at 70~K with 11 mW 857~nm excitation.}
  \label{fig:Piranha}
\end{figure}

Here, we study the surface stability after piranha cleanings. After piranha cleaning, there was no substantial change of emission statistics of \siv{} and \sivm{}, as shown in Fig.~\ref{fig:Piranha}. This suggests that surface condition and the approach to tuning SiV charge state with surface control was robust against piranha cleaning.

\subsection{Additional measurements with resonant excitation}
\begin{figure}[h!]
  \centering
  \includegraphics[width = 129mm]{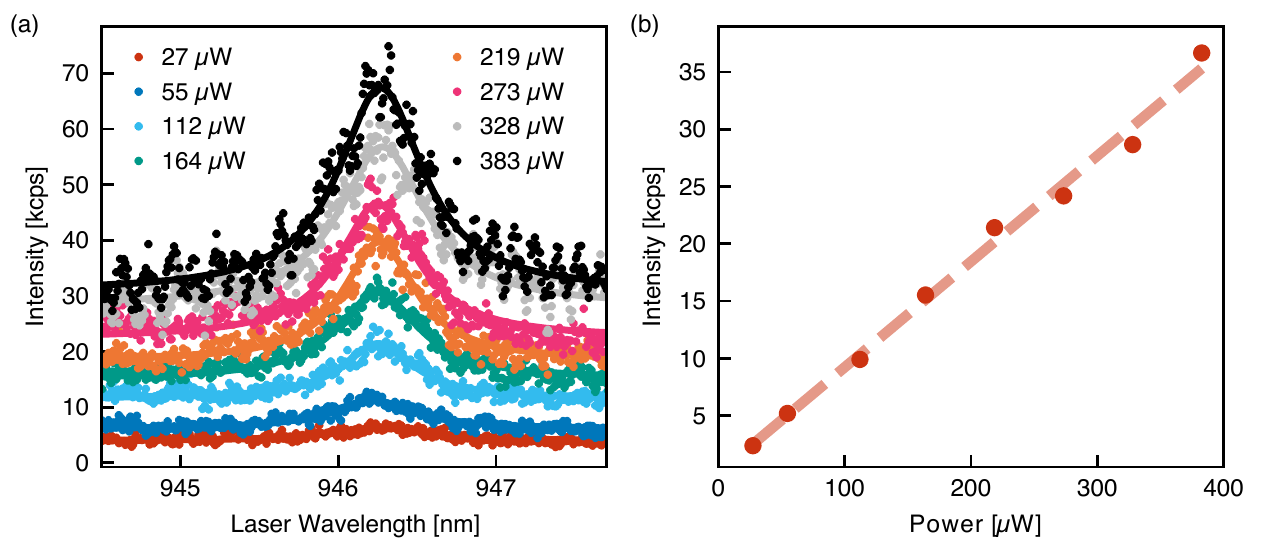}
  \caption{Power dependence of resonant excitation via the ZPL. (a) Photoluminescence excitation scans performed at different optical powers. The solid curves are Lorentzian fits to the data. (b) Peak emission intensities extracted from the fits in (a) as a function of power. The dashed line shows a fit using $I(P) = I_s/(1 + P_0/P)$ where $P$ is the optical power, $P_0$ is the saturation power and $I_s$ is the saturated emission intensity. No sign of saturation was observed. Measurements were performed at 10~K.}
  \label{fig:PLE}
\end{figure}

We performed resonant excitation via the ZPL of \siv{} in a PR with different excitation powers. The PR characterized is the same PR that was measured for ODMR in Fig.~4(c) and Fig.~4(d). The surface was first H-terminated and subsequently underwent three piranha cleanings. As shown in Fig.~\ref{fig:PLE}, no sign of saturation in intensity was observed.

\subsection{Additional ODMR measurements}
\begin{figure}[h!]
  \centering
  \includegraphics[width = 86mm]{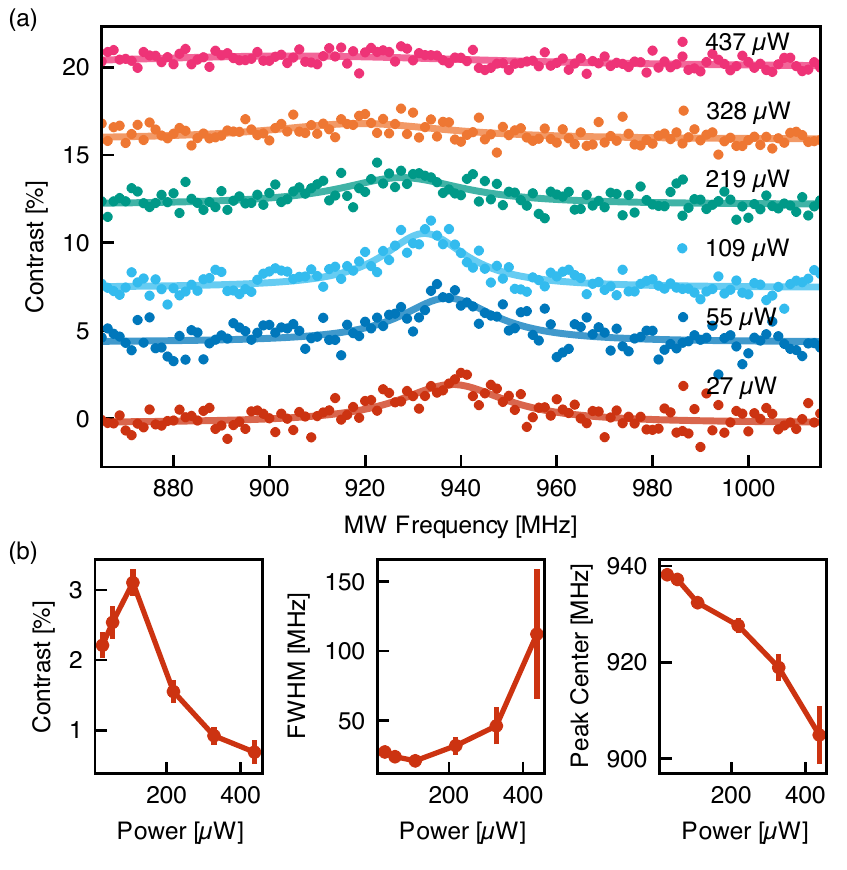}
  \caption{Optical power dependence of CW ODMR. (a) CW ODMR performed with different resonant optical powers. The laser was parked at 946.45 nm. The solid curves are Lorentzian fits to the data. (b) Fit results of the optical power dependence. Left: ODMR contrast with different optical powers. Middle: transition linewidths with different optical powers. Right: transition frequencies with different optical powers. The measurement was performed with MW power set to -23 dBm before the amplifier.}
  \label{fig:FigS2}
\end{figure}

We performed continuous-wave (CW) ODMR measurements on a PR with different optical powers (Fig.~\ref{fig:FigS2}(a)). The peak position for magnetic resonance shifts to lower frequency with higher excitation power, as shown in Fig.~\ref{fig:FigS2}(b), accompanied by linewidth broadening and contrast reduction. 

\begin{figure}[h!]
  \centering
  \includegraphics[width = 86mm]{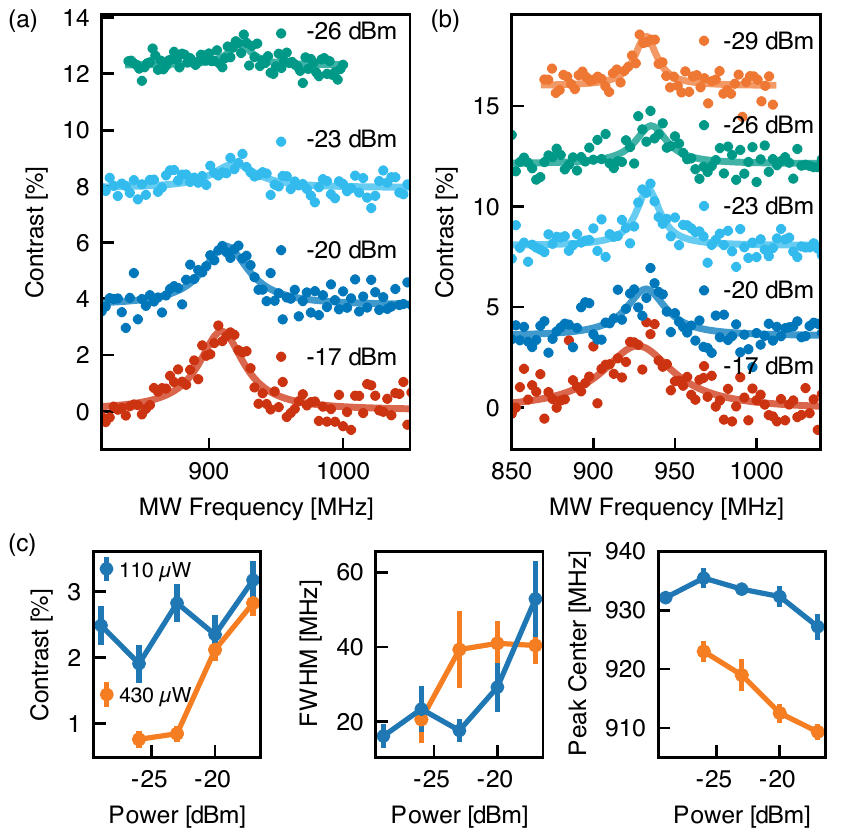}
  \caption{MW power dependence of CW ODMR. (a) CW ODMR performed with different MW powers. The laser was parked at 946.45 nm with laser power set to 430 $\mu$W. The solid curves are Lorentzian fits to the data. (b) CW ODMR performed with different MW powers. The laser was parked at 946.45 nm with laser power set to 110 $\mu$W. The solid curves are Lorentzian fits to the data. (c) Fit results of the MW power dependence of ODMR. Left: ODMR contrast with different MW powers. Middle: linewidth of ODMR with different MW powers. Right: positions of ODMR transition with different MW powers. Transition linewidths broaden at higher powers, and transition frequencies shifts to lower frequency at higher powers. The MW power quoted here is the power before the MW amplifier.}
  \label{fig:FigSMW}
\end{figure}

In addition to the optical power dependence, we also observed shift of ODMR peak position at higher MW powers. CW ODMR were measured as a function of MW power at two optical powers, as shown in Fig.~\ref{fig:FigSMW}. We observed that the ODMR position shifted to lower frequency at higher MW power, and the MW power dependence was more significant for higher optical power (Fig.~\ref{fig:FigSMW}(c)). The origin of the power dependent shift is subject to further investigation.

\section{Estimation of the number of SiV centers inside parabolic reflectors}

Here, we estimate the average number of SiV centers inside each PR based on the occurrence of \siv{} and \sivm{} centers measured from a large number (324) of PRs. As shown in Fig.~\ref{fig:Piranha}(d), we observed a maximum occurrence of 82.7\%  for \siv{} under H-termination with one piranha cleaning and a maximum occurrence of 93.5\% for \sivm{} under O-termination. Assuming the number of SiV centers in a PR follows a Poisson distribution $P(n,\lambda)$, where $n$ is the number of SiV centers and $\lambda$ is the average number of SiV centers per PR, the occurrence of SiV centers is therefore
\begin{equation}
    P(n> 0) = 1 - P(0,\lambda)
\end{equation}
Using the occurrence extracted from the measured spectral statistics, we estimate the average number of SiV centers per PR to be $1.8 \pm 0.2$ for \siv{} and $2.7 \pm 0.4$ for \sivm{} (95 \% confidence interval). Based on the PR dimension (300~nm diameter) and the implantation dose (3$\times$ $10^{11}$ cm$^{-2}$) used in our sample, the maximum number of silicon ions per PR is 212. The apparent occurrence here is similar to prior work on NV centers in PRs after accounting for differences in the implantation dose and the dimension of PR \cite{Hedrich2020}. 

We note that this estimation is a conservative lower bound because using the occurrence alone to estimate the mean of a Poisson distribution is sensitive to experimental variation. Imperfection of PR fabrication and the inability to observe SiV centers with emission intensity below the background level would result in a lower apparent occurrence. For example, even with only one PR showing no SiV emission (99.7\% occurrence with 324 sampling), this estimation already limits the mean number of SiV centers per PR to $5.8 \pm 2.0$.

To properly estimate how many SiV centers are formed in each PR under based on the emission intensity distribution, one would need some prior knowledge of the emission statistics of SiV centers such as the average brightness and the variation of brightness. Alternatively, if the average brightness of individual SiV center is much higher than the variation their brightness, we would observe isolated peaks in the intensity distribution and the position of the peaks would be related to the number of SiV centers. However, we observe no such peaks in our histogram (Fig.~\ref{fig:Piranha}(a), Fig.~\ref{fig:Piranha}(b)), suggesting the variation of SiV brightness in the PRs is at least comparable to the intensity of individual SiV centers, making it challenging to provide a direct estimation of the average number of SiV inside each PR.

\section{Unknown defect with 992 nm emission}
\begin{figure}[h!]
  \centering
  \includegraphics[width = 86mm]{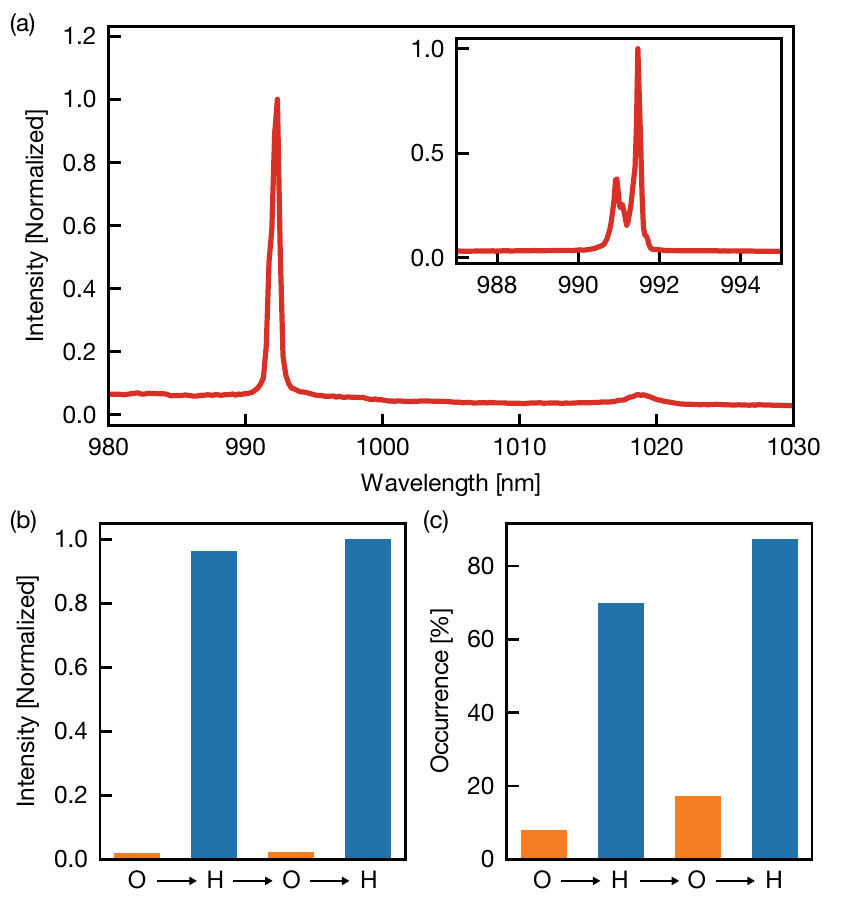}
  \caption{Unknown defect at 992 nm. (a) Emission spectrum measured on the high purity sample under H-termination. Inset: spectrum with higher resolution showing a doublet structure. (b) Average intensity of 992~nm emission under different surfaces. (c) Occurrence of 992~nm emission under different surfaces. The surface was toggled between O-termination (orange) and H-termination (blue). Measurements were performed with $\sim$11~mW 857 nm excitation at 10~K.}
  \label{fig:Fig992}
\end{figure}

Throughout our measurement, we observed an additional emission feature around 992 nm. The 992 nm emission correlates with another broad feature at 1020 nm (Fig.~\ref{fig:Fig992}(a)). We observed that the 992 nm emission was also dependent on the surface termination, where it was prominent under H-termination and was suppressed under O-termination, as shown in Fig.~\ref{fig:Fig992}(b) and Fig.~\ref{fig:Fig992}(c).

The origin of this additional emission feature remains unknown; the observation of 992~nm emission in the unimplanted surface appears to rule out a silicon-related defect. Nevertheless, the correlation between this defect and the surface termination demonstrates that our method of tuning the Fermi level via surface control should be applicable to a broad range of defects in diamond.

\clearpage
\bibliography{ChemicalControl}